\documentclass{aa}  

\usepackage{graphicx}
\usepackage{txfonts}
\usepackage{amssymb}
\usepackage{natbib}
\bibpunct{(}{)}{;}{a}{}{,}

\def \cmtwo{cm$^{-2}$}
\def \cmthree{cm$^{-3}$}

\def \kms{km~s$^{-1}$}
\def \ss{s$^{-1}$}
\def \Kkms{K~km~s$^{-1}$}

\def \oofull{\mbox{O$_2$~3$_{3}$--1$_{2}$}}

\def \thco{$^{13}$CO}
\def \ceio{C$^{18}$O}

\def \hh{H$_2$}
\def \hho{H$_2$O}

\def \oo{O$_2$}

\def \co{CO}
\def \no{NO}
\def \xoo{$X$(O$_{2}$)}

\def \rms{$rms$}
\def \ncr{$n_{\rm cr}$}
\def \Eup{$E_{\rm up}$}
\def \Tpeak{$T_{\rm peak}$}
\def \TmbdV{$\int T_{\rm mb} {\rm d}V$}

\def \G0{$G_{0}$}
\def \vlsr{$V_{\rm LSR}$}        

\def \NOO{$N$(O$_2$)}

\begin{document} 

   \title{Deep observations of O$_{2}$ toward a low-mass protostar with  \textit{Herschel}-HIFI\thanks{{\it Herschel} is an ESA space observatory with science instruments provided by European-led Principal Investigator consortia and with important participation from NASA.}}

  \authorrunning{Y{\i}ld{\i}z et al.}
  \titlerunning{Deep observation of O$_{2}$ toward the NGC~1333~IRAS~4A protostar}

\author{
     Umut~A.~Y{\i}ld{\i}z\inst{\ref{inst1}},
     Kinsuk~Acharyya\inst{\ref{inst2}},
     Paul~F.~Goldsmith\inst{\ref{inst3}},
     Ewine~F.~van~Dishoeck\inst{\ref{inst1},\ref{inst4}},
     Gary~Melnick\inst{\ref{inst5}},
	 Ronald Snell\inst{\ref{inst6}}
     Ren\'{e}~Liseau\inst{\ref{inst7}},
     Jo-Hsin~Chen\inst{\ref{inst3}},
     Laurent~Pagani\inst{\ref{inst8}},
     Edwin Bergin\inst{\ref{inst9}},
     Paola Caselli\inst{\ref{inst10},\ref{inst11}},
     Eric Herbst\inst{\ref{inst12}},
     Lars~E.~Kristensen\inst{\ref{inst5}},
     Ruud~Visser\inst{\ref{inst9}},
	 Dariusz~C.~Lis\inst{\ref{inst13}}, 
     Maryvonne Gerin\inst{\ref{inst14}}
}

\institute{
Leiden Observatory, Leiden University, PO Box 9513, 2300 RA Leiden, The Netherlands\label{inst1}\thanks{\email{yildiz@strw.leidenuniv.nl}}
\and
S.N. Bose National Centre for Basic Sciences, Salt Lake, Kolkata, 700098, India\label{inst2}
\and
Jet Propulsion Laboratory, California Institute of Technology, 4800 Oak Grove Drive, Pasadena CA, 91109, USA\label{inst3}
\and
Max Planck Institut f\"{u}r Extraterrestrische Physik, Giessenbachstrasse 1, 85748 Garching, Germany\label{inst4}
\and
Harvard-Smithsonian Center for Astrophysics, 60 Garden Street, Cambridge, MA 02138, USA\label{inst5}
\and
Department of Astronomy, LGRT 619, University of Massachusetts, 710 North Pleasant Street, Amherst, MA 01003, USA\label{inst6}
\and
Dept. of Earth \& Space Sciences, Chalmers University of Technology, Onsala Space Observatory, SE-439 92 Onsala, Sweden\label{inst7}
\and
LERMA \& UMR8112 du CNRS, Observatoire de Paris, 61 Av. de l'Observatoire, 75014, Paris, France\label{inst8}
\and
Department of Astronomy, University of Michigan, 500 Church Street, Ann Arbor, MI 48109-1042, USA\label{inst9}
\and
School of Physics and Astronomy, University of Leeds, Leeds LS2 9JT, UK\label{inst10}
\and
INAF-Osservatorio Astrofisico di Arcetri, Largo E. Fermi 5, I-50125 Firenze, Italy\label{inst11}
\and
Departments of Chemistry, Astronomy, and Physics, The University of Virginia, Charlottesville, Virginia, USA\label{inst12}
\and
California Institute of Technology, Cahill Center for Astronomy and Astrophysics 301-17, Pasadena, CA 91125, USA\label{inst13}
\and
LRA/LERMA, CNRS, UMR8112, Observatoire de Paris \& {\'E}cole Normale Sup{\'e}rieure, 24 rue Lhomond, 75231 Paris Cedex 05, France\label{inst14}
}

   \date{Draft: xxxx xx xx}

\def\placeTableOverviewOfTheObservations{
\begin{table}[!t]
\caption{Overview of the observed lines.}
\tiny
\begin{center}
\begin{tabular}{l c r r r r r r r r r r}
\hline \hline
Molecule & Transition & $E_\mathrm{u}/k_{\mathrm{B}}$ & $A_{\rm ul}$ & Frequency \\
 &   &  [K] & [\ss] & [GHz]  \\
\hline
\noalign{\smallskip}
O$_2$     & $N_{J}$=3$_{3}$--1$_{2}$ & 26.4   & 8.657$\times$10$^{-9}$   & 487.2492640 \\
C$^{18}$O & $J$=1--0                  & \phantom{1}5.3   & 6.266$\times$10$^{-8}$ & 109.7821734   \\
C$^{18}$O & $J$=3--2                  & 31.6 & 2.172$\times$10$^{-6}$     & 329.3305525           \\
C$^{18}$O & $J$=5--4                  & 79.0 & 1.062$\times$10$^{-5}$     & 548.8310055     \\
NO (1) & $J$=5/2--3/2,\,$F$=3/2--1/2 & 19.3   & 1.387$\times$10$^{-6}$   & 250.8169540             \\
NO (2) & $J$=5/2--3/2,\,$F$=5/2--3/2 & 19.3   & 1.553$\times$10$^{-6}$   & 250.8155940             \\
NO (3) & $J$=5/2--3/2,\,$F$=7/2--5/2 & 19.3   & 1.849$\times$10$^{-6}$   & 250.7964360             \\
NO (4) & $J$=5/2--3/2,\,$F$=3/2--3/2 & 19.3   & 4.437$\times$10$^{-7}$   & 250.7531400             \\
\noalign{\smallskip}
\hline
\end{tabular}
\end{center}
\label{tbl:overviewobs}
\end{table}
}

\def\placeTableOverviewOfTheObservationResults{
\begin{table*}[!t]
\caption{Summary of the observed line intensities in a 44$\arcsec$ beam.}
\small
\begin{center}
\begin{tabular}{l r r | r r r | r r r | r r r}
\hline \hline
Molecule & Transition & Telescope/ & $\int T_{\mathrm{MB}} \mathrm{d}V$ & $T_{\mathrm{peak}}$ & FWHM & $\int T_{\mathrm{MB}} \mathrm{d}V$ & $T_{\mathrm{peak}}$ & FWHM & $rms$\\
 & & Instrument & [\Kkms] & [K]  & [\kms] &  [\Kkms] & [K]  & [\kms] & [mK] \\
 &   &  &    \multicolumn{3}{c}  {7.0 \kms\ component\tablefootmark{a}} &   \multicolumn{3}{c}  {8.0 \kms\ component\tablefootmark{b}} \\%
\hline
\noalign{\smallskip}
O$_2$ & $N_{J}$=3$_{3}$--1$_{2}$ & {\it Herschel}-HIFI &  $<0.0027$\tablefootmark{c}           &\dots\phantom{0} & \dots & 0.0069  & 0.0046 & 1.3     & 1.3\tablefootmark{d} \\
C$^{18}$O & $J$=1--0          & IRAM~30m-EMIR       & 1.30\phantom{0}\phantom{0}\phantom{0} & 1.38\phantom{0} & 0.9   & 2.25\phantom{0}\phantom{0} & 2.35\phantom{0}\phantom{0} & 0.9  & 26\tablefootmark{e} \\
C$^{18}$O & $J$=3--2          & JCMT-HARP-B         & 1.32\phantom{0}\phantom{0}\phantom{0} & 1.36\phantom{0} & 0.9   & 1.67\phantom{0}\phantom{0} & 1.74\phantom{0}\phantom{0} & 0.9  & 99\tablefootmark{e} \\
C$^{18}$O & $J$=5--4          & {\it Herschel}-HIFI & 0.39\phantom{0}\phantom{0}\phantom{0} & 0.36\phantom{0} & 1.0   & 0.13\phantom{0}\phantom{0} & 0.13\phantom{0}\phantom{0} & 1.0  & 10\tablefootmark{e} \\
NO (3) & $J$=5/2--3/2,\,$F$=7/2--5/2 & JCMT-RxA            & $<$0.15\tablefootmark{c}\phantom{0}\phantom{0}  & \dots\phantom{0}\phantom{0} & \dots & 0.18\phantom{0}\phantom{0}& 0.16\phantom{0}\phantom{0} & 2.9  & 46\tablefootmark{e} \\
\noalign{\smallskip}
\hline
\end{tabular}
\end{center}
\tablefoot{The values are calculated through a fit to the lines.
\tablefoottext{a}{\vlsr=7.0~\kms\ component.}
\tablefoottext{b}{\vlsr=8.0~\kms\ component.}
\tablefoottext{c}{3$\sigma$ upper limit.}
\tablefoottext{d}{In 0.35~\kms\ bins.}
\tablefoottext{e}{In 0.3~\kms\ bins.}
}
\label{tbl:overviewobsresults}
\end{table*}
}


 
  \abstract
   {According to traditional gas-phase chemical models,
  \oo\ should be abundant in molecular clouds, but until recently, attempts to
  detect interstellar \oo\ line emission with ground- and
  space-based observatories have failed.}
   {Following the multi-line detections of \oo\ with low abundances in the 
   Orion and $\rho$~Oph~A molecular clouds with {\it Herschel}, it is important 
   to investigate other environments, and we here quantify the \oo\ abundance 
   near a solar-mass protostar.}
   {Observations of molecular oxygen,
   \oo, at 487~GHz toward a deeply embedded low-mass
   Class~0 protostar, \object{NGC~1333-IRAS~4A}, are presented, using
   the Heterodyne Instrument for the Far Infrared (HIFI) on the
   \textit{Herschel} Space Observatory. Complementary data
   of the chemically related NO and CO molecules are obtained as well.
  The high spectral
    resolution data are analysed using radiative transfer
     models to infer column densities and abundances, and are tested
  directly  against full gas-grain chemical models.}
{ The deep HIFI spectrum fails to show \oo\ at the velocity of the
  dense protostellar envelope, implying one of the lowest abundance upper
  limits of O$_2$/H$_2$ at $\leq$6$\times$10$^{-9}$ (3$\sigma$). The
  O$_2$/CO abundance ratio is less than 0.005. However, a tentative
  (4.5$\sigma$) detection of \oo\ is seen at the velocity of the
  surrounding NGC~1333 molecular cloud, shifted by 1 km s$^{-1}$ 
  relative to the protostar. 
  For the protostellar envelope, pure gas-phase models and gas-grain chemical models require
  a long pre-collapse phase ($\sim$0.7--1$\times$10$^{6}$~years), 
  during which atomic and molecular oxygen are frozen out onto dust 
  grains and fully converted to \hho, to avoid overproduction of \oo\ 
  in the dense envelope.
  The same model also reproduces the limits on the
  chemically related NO molecule if hydrogenation of NO on the grains to more
  complex molecules such as NH$_2$OH, found in recent laboratory
  experiments, is included. The tentative detection of O$_2$ in the
  surrounding cloud is consistent with a low-density PDR model with
  small changes in reaction rates.}
   {The low \oo\ abundance in the collapsing envelope around a low-mass
 protostar suggests that the gas and ice entering protoplanetary disks is very poor in \oo.}


   \maketitle
%


\section{Introduction}
Even though molecular oxygen (\oo) has a simple chemical structure, it
remains difficult to detect in the interstellar medium after many
years of searches \citep[][and references therein]{Goldsmith11}.  Oxygen
is the third most abundant element in the Universe, after hydrogen and
helium, which makes it very important in terms of understanding the
formation and evolution of the chemistry in astronomical sources.

Pure gas-phase chemistry models suggest a steady-state abundance of
\xoo$\approx$7$\times$10$^{-5}$ relative to \hh\
\citep[e.g., Table 9 of][]{Woodall07}, however observations show that the
abundance is several orders of magnitude lower than these model
predictions.  Early (unsuccessful) ground-based searches of O$_2$ were
done through the $^{16}$O$^{18}$O isotopologue
\citep{Goldsmith85,Pagani93}, for which some of the lines fall in a transparent
part of the atmosphere.
Due to the oxygen content of the Earth's atmosphere, it is
however best to observe \oo\ from space.
Two previous space missions, the {\it Submillimeter Wave Astronomy
Satellite} \citep[\textit{SWAS;}][]{Melnick00} and the {\it Odin
Satellite} \citep{Nordh03} were aimed at detecting and studying 
interstellar molecular oxygen through specific transitions. {\it SWAS}
failed to obtain a definitive detection of \oo\ at 487~GHz toward
nearby clouds \citep{Goldsmith00}, whereas {\it Odin} observations of
\oo\ at 119~GHz gave upper limits of $\leq$10$^{-7}$
\citep{Pagani03}, except for the $\rho$~Ophiuchi A cloud
\citep[\xoo$\sim$5$\times$10$^{-8}$;][]{Larsson07}.

The {\it Herschel} Space Observatory provides much higher spatial
resolution and sensitivity than previous missions and therefore allows
very deep searches for O$_2$. Recently, {\it Herschel}-HIFI provided
firm multi-line detections of \oo\ in the Orion and $\rho$~Oph~A
molecular clouds \citep{Goldsmith11, Liseau2012}. The abundance was
found to range from \xoo$\simeq$10$^{-6}$ (in Orion) to
\xoo$\simeq$5$\times$10$^{-8}$ (in $\rho$~Oph~A). The interpretation
of the low abundance is that oxygen atoms are frozen out onto grains
and transformed into water ice that coats interstellar dust, leaving
little atomic O in the gas to produce \oo\ \citep{Bergin00}. So far,
\oo\ has only been found in clouds where (external) starlight has
heated the dust and prevented atomic O from sticking onto the grains
and being processed into \hho\, as predicted by theory \citep{Hollenbach09} or 
where \oo\ is enhanced in postshock gas \citep{Goldsmith11}.  Not every warm
cloud has O$_2$, however. \citet{Melnick12} report a low upper limit
on gaseous \oo\ toward the dense Orion Bar photon-dominated region
(PDR).

\begin{figure}[!t]
\begin{center}
\includegraphics[scale=0.45]{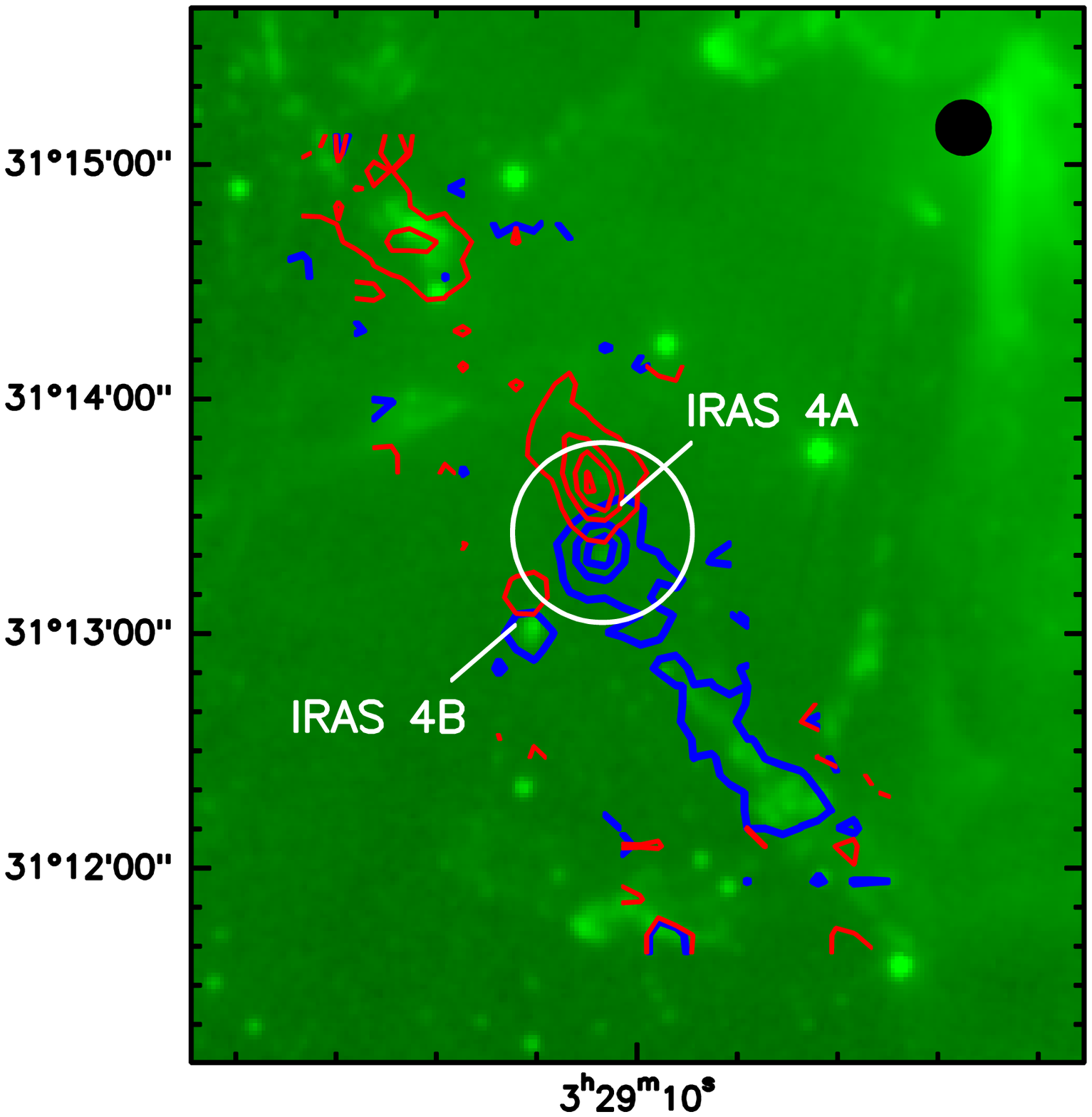}
\end{center}
\caption{\small{{\it Spitzer}/IRAC1 \citep{Gutermuth08} and CO~6--5
contours \citep{Yildiz12} are overlaid to illustrate the NGC 1333
IRAS 4A and 4B protostars. The white circle in the center
represents the observed HIFI beam of 44$\arcsec$ centred on IRAS
4A, illustrating that it partially overlaps with the outer part of
the IRAS 4B envelope. The contours indicate the outflows, with
levels starting from 15~\Kkms\ with an increasing step
size of 30~\Kkms. Blue and red velocity ranges
are selected from --20 to 2.7 and from 10.5 to 30~\kms,
respectively. The black dot on upper right corner shows the beam
size of the CO~6--5 data. }}
\label{fig:iras4abSpitzerCO}
\end{figure}

Although the detection of \oo\ in some molecular clouds is
significant, these data tell little about the presence of \oo\ in
regions where solar systems may form. It is therefore important to
also make deep searches for \oo\ near solar-mass protostars to
understand the origin of molecular oxygen in protoplanetary disks and
eventually (exo-)planetary atmospheres. 
Even though the bulk of the \oo\ in the Earth's atmosphere arises from microorganisms,
the amount of \oo\ that could be delivered by cometesimal impacts needs to be quantified.
In the
present paper, a nearby low-mass deeply embedded protostar, NGC~1333 IRAS~4A, is
targeted, which has one of the highest line of sight hydrogen column
densities of $N$(\hh) $\sim$10$^{24}$~\cmtwo\ derived from dust modeling
\citep{Jorgensen02,Kristensen12}. Since the {\it Herschel} beam size at
487~GHz is a factor of $\sim$6 smaller than that of {\it SWAS}, 
{\it Herschel} is much more sensitive to emission from these compact sources.
Protostars also differ from dense clouds or PDRs by the fact
that a significant fraction of the dust is heated internally by the
protostellar luminosity to temperatures above those needed to
sublimate O and \oo.

\object{NGC~1333~IRAS~4A} is located in the southeast part of the
NGC~1333 region, together with IRAS~4B (henceforth \object{IRAS~4A}
and \object{IRAS~4B}).  A distance of 235$\pm$18~pc is adopted based
on VLBI parallax measurements of water masers in the nearby source
SVS~13 \citep{Hirota08}.  Both objects are classified as
deeply-embedded Class~0 low-mass protostars \citep{Andre94} and are
well-studied in different molecular lines such as CO, SiO, H$_2$O and
CH$_3$OH
\citep[e.g.,][]{Blake95,Lefloch98_SiO,Bottinelli07,Yildiz12,Kristensen10hifi}.
Figure \ref{fig:iras4abSpitzerCO} shows a CO $J$=6--5 contour map
obtained with APEX \citep{Yildiz12} overlaid on a {\it
Spitzer}/IRAC1 (3.6~$\mu{\rm m}$) image \citep{Gutermuth08}. Both
IRAS~4A and IRAS~4B have high-velocity outflows seen at different
inclinations.
The projected separation between the centers of IRAS~4A and
IRAS~4B is 31$\arcsec$ ($\sim$7300 AU).  The source IRAS4A was
chosen for the deep \oo\ search because of its chemical richness and
high total column density. In contrast to many high-mass
protostars, it has the advantage that even 
very sensitive spectra do not show line confusion.
  
On a larger scale, early millimeter observations of CO and $^{13}$CO $J$=1--0
by \citet{Loren76} and \citet{Liseau88} found two (possibly colliding) clouds in
the NGC~1333 region, with velocities separated by up to 2 \kms.
\citet{Cernis1990} used extinction mapping in the NGC~1333 region
to confirm the existence of two different clouds. The IRAS 4A
protostellar envelope is centred at the lower velocity around
\vlsr=7.0 \kms, whereas the lower (column) density cloud appears
around \vlsr=8.0 \kms. The high spectral resolution of our data allows
O$_2$ to be probed in both clouds.  Optically thin isotopologue data
of C$^{18}$O~$J$=1--0 up to $J$=5--4 are used to characterize the conditions
in the two components.
Note that these velocities do not overlap with those of the red outflow lobe,
which start at \vlsr=+10.5~\kms.

We present here the first sensitive observations of the \oofull\ 487~GHz
line towards a deeply embedded low-mass Class~0 protostar, observed
with {\it Herschel}-HIFI. Under a wide range of conditions, the \oo\
line at 487~GHz is the strongest, therefore this line is selected
for long integration. The data are complemented by ground-based
observations of CO isotopologues and NO using the IRAM~30m and JCMT
telescopes. The CO data are used to characterize the kinematics and
physical conditions in the clouds as well as the column of gas where CO is not
frozen out. Since the \oo\ ice has a very similar binding energy as the CO
ice, either in pure or mixed form \citep{Collings04,Acharyya07}, CO
provides a good reference for \oo. NO is chosen because it is a
related species that could help to constrain the chemistry of \oo. In the gas, \oo\ can be produced from atomic O through the reaction
\citep{HerKlemp73,BlackSmith84}
\begin{equation}
{\rm O + OH \to O_2 + H}
\label{eq:o2reaction}
\end{equation}
with rate constants measured by \citet{Carty06}.
The nitrogen equivalent of Eq.~(\ref{eq:o2reaction}) produces NO through
\begin{equation}
{\rm N + OH \to NO + H.}
\label{eq:noreaction}
\end{equation}

\placeTableOverviewOfTheObservations

\begin{figure*}[!t]
\begin{center}
\includegraphics[scale=0.65]{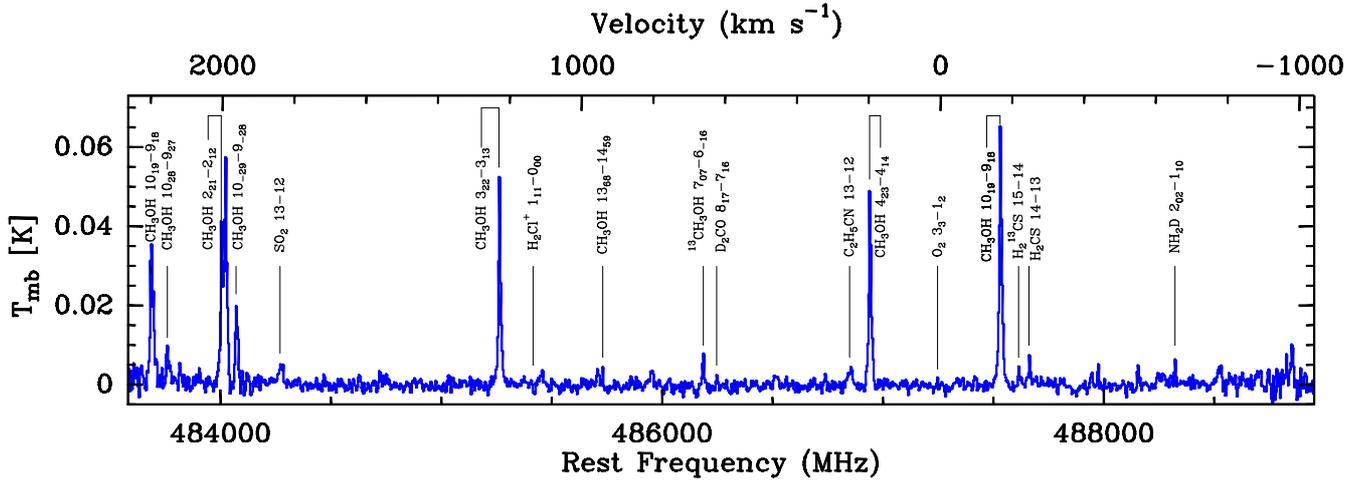}
\end{center}
\caption{\small{Full spectrum taken with HIFI, with the H- and
    V-polarization spectra averaged. The frequency range is 483.59~GHz
    to  488.94~GHz from left to right. The entire bandwidth is
    5.35~GHz.  The \oo\ line is centred near \vlsr=7.0~\kms. 
    A blow-up of the spectrum is presented in Fig.~A.1.}}
\label{fig:iras4a487ghzfull}
\end{figure*}

The outline of the paper is as follows. Section~\ref{sec:observations}
describes the observations and the telescopes where the data were
obtained.  Results from the observations are presented in
Section~\ref{sec:results}.  The deep HIFI spectrum reveals a
non-detection of \oo\ at the velocity of the central protostellar
source. However, a tentative (4.5$\sigma$) detection is found
originating from the surrounding NGC~1333 cloud at \vlsr=8 \kms. In
Section 3, a physical model of the source coupled with line
radiative transfer is used to infer the gas-phase
abundance profiles of CO, \oo\ and NO in the protostellar envelope
that are consistent with the data (`backward' or retrieval
modeling, see \citealt{Doty04} for terminology). Forward modeling
using a full gas-grain chemical code coupled with the same
physical model is subsequently conducted to interpret the
non-detection in Section~\ref{sec:modeling}.  In
Section~\ref{sec:tentative}, the implications for the possible
detection in the 8 km s$^{-1}$ component are discussed and in
Section~\ref{sec:conclusions}, the conclusions from this work are
summarized.


\section{Observations}
\label{sec:observations}

\placeTableOverviewOfTheObservationResults

The molecular lines observed towards the IRAS~4A protostar
\citep[$3^{\mathrm{h}}29^{\mathrm{m}}10\fs5$,
  $+31\degr13\arcmin30\farcs9$ (J2000);][]{Jorgensen09} are presented
in Table~\ref{tbl:overviewobs} with the corresponding frequencies,
upper level energies ($E_\mathrm{u}/k_{\mathrm{B}}$), and Einstein $A$
coefficients.  The O$_2$ data were obtained with the Heterodyne
Instrument for the Far-Infrared \citep[HIFI;][]{degraauw10} onboard the
\textit{Herschel} Space Observatory \citep{Pilbratt10}, in the context
of the `{\it Herschel} Oxygen Project' (HOP) open-time key program,
which aims to search for \oo\ in a range of star-forming regions and
dense clouds \citep{Goldsmith11}. Single pointing observations at the
source position were carried out on operation day OD~445 on August~1
and 2, 2010 with {\it Herschel} obsids of
1342202025-$\dots$-1342202032. The data were taken in
dual-beam switch (DBS) mode using the HIFI band 1a mixer with a chop reference
position located 3$\arcmin$ from the source position.  Eight
observations were conducted with an integration time of 3477 seconds
each, and eight different local-oscillator (LO) tunings were used in
order to allow deconvolution of the signal from the image side band. 
The LO tunings are shifted by 118~MHz up to 249~MHz. 
Inspection of the data shows no contamination from the
reference position in any of the observations, nor from the image
side-band. The total integration time is thus 7.7 hours (27816
seconds) for the on+off source integration.

The central frequency of the \oofull\ line is 487.249264~GHz with an
upper level energy of $E_{\rm u}$=26.4~K and an Einstein $A$
coefficient of 8.657$\times$10$^{-9}$~s$^{-1}$ \citep{Drouin10}.  In
HIFI, two spectrometers are in operation, the ``Wide Band
Spectrometer'' (WBS) and the ``High Resolution Specrometer'' (HRS)
with resolutions of 0.31 \kms\ and 0.073 \kms\ at 487~GHz,
respectively.  Owing to the higher noise ranging from a factor of 1.7
up to 4.7 of the HRS compared with the WBS, only WBS observations were
used in the analysis.  There is a slight difference between the pointings
of the H and V polarizations in HIFI, but this difference of $\Delta
HV$ \citep[--6$\farcs$2, +2$\farcs$2;][]{Roelfsema11} for Band
1 is small enough to be neglected relative to the beam size of 44$\arcsec$ (FWHM). Spectra from both polarizations
were carefully checked for differences in intensities of other
detected lines but none were found. Therefore the two polarizations
were averaged to improve the signal to noise ratio.

Data processing started from the standard HIFI pipeline in the
\textit{Herschel} Interactive Processing Environment
(HIPE\footnote{HIPE is a joint development by the Herschel Science
  Ground Segment Consortium, consisting of ESA, the NASA Herschel
  Science Center, and the HIFI, PACS and SPIRE consortia.}) ver.~8.2.1
\citep{OttS10}, where the \vlsr\ precision is of the order of a few
m~s$^{-1}$.  The lines suffer from significant standing waves in each of
the observations. Therefore a special task \verb1FitHifiFringe1 in
HIPE was used to remove standing waves. The fitting routine was
applied to each observation one by one and it successfully removed a
large part of the standing waves.  Further processing and analysis was
done using the
GILDAS-\verb1CLASS1\footnote{{http://www.iram.fr/IRAMFR/GILDAS/}}
software. A first order polynomial was applied to all observations, 
which were subsequently averaged together. The standard
antenna temperature scale $T_{\rm A}^{*}$ is corrected to the main
beam temperature $T_{\rm MB}$ \citep{KutnerUlich81} by applying the 
efficiency of 0.76 for HIFI band 1a 
\citep[][Fig.~\ref{fig:iras4a487ghzfull}]{Roelfsema11}.

To understand and constrain the excitation and chemistry of \oo, complementary 
transitions in NO and \ceio\ were observed.
Nitrogen monoxide (NO) was observed with the James Clerk Maxwell
Telescope (JCMT\footnote{The JCMT is operated by the Joint Astronomy
Centre on behalf of the Science and Technology Facilities Council of
the United Kingdom, the Netherlands Organisation for Scientific
Research, and the National Research Council of Canada.}) by using
Receiver A with a beam size of 20$\arcsec$ as part of the M10BN05 
observing program.  The total integration time for this observation was 91 minutes.
\mbox{C$^{18}$O~$J$=1--0} was observed with the IRAM~30m telescope\footnote{Based on
observations carried out with the IRAM 30m Telescope. IRAM is
supported by INSU/CNRS (France), MPG (Germany) and IGN (Spain).}
using a frequency-switch mode over an area of 
1$\arcmin$$\times$1$\arcmin$ in a 22$\arcsec$ beam.  
A \mbox{C$^{18}$O~$J$=3--2} spectrum was extracted from the large NGC~1333 map of 
\citet{Curtis10_1data}, which was observed with the HARPB instrument at 
JCMT with position switch-mode 
(off position coordinate: 
$3^{\mathrm{h}}29^{\mathrm{m}}00\fs0$, $+31\degr52\arcmin30\farcs0$; J2000) 
in a 15$\arcsec$ beam \citep[also in][]{Yildiz12}.
Both maps were convolved to a beam of 44$\arcsec$ in order to directly compare
with the \oo\ spectra in the same beam. 
The 15$\arcsec$ beam spectra presented in \citet{Yildiz12} 
show primarily the 7.0 \kms\ component.
The \mbox{C$^{18}$O~$J$=5--4} line was observed with {\it Herschel}-HIFI 
within the ``Water in Star-forming regions with {\it Herschel}'' (WISH) 
guaranteed-time key program \citep{vanDishoeck11} in a beam size of 
40$\arcsec$ and reported in \citet{Yildiz12}.
Beam efficiencies are 0.77, 0.63, and 0.76 for the 1--0,
3--2, and 5--4 lines, respectively.  The calibration uncertainty for
HIFI band 1a is 15\%, whereas it is 20\% for the IRAM~30m and JCMT
lines.

The HIFI beam size at 487~GHz of $\sim$44$\arcsec$ corresponds to a 
5170~AU radius for IRAS~4A at 235~pc
(Fig.~\ref{fig:iras4abSpitzerCO}, white circle).  It therefore
overlaps slightly with the dense envelope around IRAS~4B \citep[see
also Fig.~13 in][]{Yildiz12} but this is neglected in the
analysis. The NO data were taken as a single pointing observation,
therefore the beam size is $\sim$20$\arcsec$, about half of the
diameter covered with the \oo\ observation. 



\section{Results}
\label{sec:results}
In Fig.~\ref{fig:iras4a487ghzfull}, the full {\it Herschel}-HIFI WBS
spectrum is presented. Although the bandwidth of the WBS data is 4~GHz,
the entire spectrum covers 5.35~GHz as a result of combining eight
different observations where the LO frequencies were slightly shifted
in each of the settings. The \rms\ of this spectrum is 1.3~mK in
0.35~\kms\ bin, therefore many faint lines are detected near
the main targeted \oofull\ line. These lines include some methanol
(CH$_{3}$OH) lines, together with e.g., SO$_{2}$, NH$_2$D, and
D$_2$CO lines.
These lines are shown in Fig.~\ref{fig:iras4a487ghzpanels} (in the
Appendix) in detail, and are tabulated with the observed information
in Table~\ref{tbl:overviewobsHIFI} (in the Appendix).

\begin{figure}[!t]
\begin{center}
\includegraphics[scale=0.37]{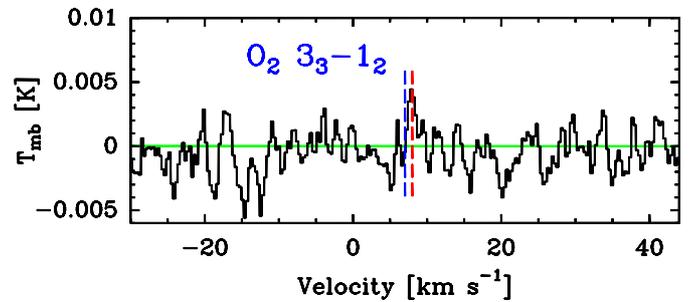}
\end{center}
\caption{\small{Spectrum of Fig.~\ref{fig:iras4a487ghzfull} magnified
around the \oofull\ line. The blue dashed line indicates the LSR velocity of the 
IRAS~4A envelope at 7.0~\kms and the red dashed line shows the velocity at 8.0~\kms.}}
\label{fig:O2only}
\end{figure}

\begin{figure}[!t]
\begin{center}
\includegraphics[scale=0.37]{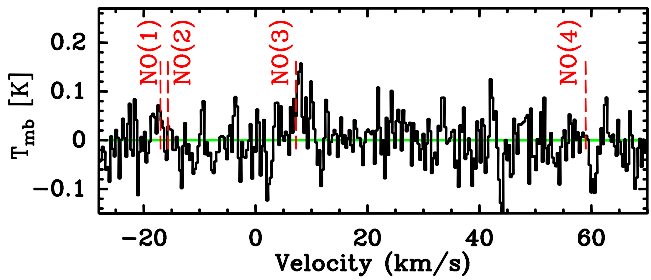}
\end{center}
\caption{\small{Spectrum of the NO $J$=5/2--3/2 transitions showing the 
location of four hyperfine (HF) components, where the details of the lines are 
given in Table~\ref{tbl:overviewobs}. The spectrum is centred on 
the NO (3) (HF) component.
    }}
\label{fig:nospectrum}
\end{figure}

\subsection{O$_2$}

A blow-up of the HIFI spectrum centred around the \mbox{\oo\ $N_{J}$=3$_{3}$--1$_{2}$} at 487~GHz
position is presented in Fig.~\ref{fig:O2only}. The source velocity of
IRAS~4A is \vlsr=7.0~\kms\ as determined from many \ceio\ lines
\citep{Yildiz12}, and is indicated by the blue dashed line in the
figure.  This spectrum of 7.7 hours integration time staring at the
IRAS~4A source position is still not sufficient for a firm detection
of the \oo\ line at 487 GHz at the source velocity. However, a
tentative detection at \vlsr=8.0~\kms\ (red dashed line in 
Fig.~\ref{fig:O2only}) is seen and will be discussed
in more detail in Sect.~\ref{sec:tentative}.

\subsection{NO}
In Fig.~\ref{fig:nospectrum}, the JCMT spectrum covering the hyperfine components
of the NO $J$=5/2--3/2 transitions are presented. For this specific transition,
the expected ratios of the line intensities in the optically thin
limit are NO~(1)~:~NO~(2)~:~NO~(3)~:~NO~(4)~=~75:126:200:24. The JCMT
observations have an \rms\ of 46~mK in 0.3~\kms\ bin and 4$\sigma$ emission is detected
only at the intrinsically strongest hyperfine transition, NO~(3), with
an integrated intensity of 0.18~\Kkms\ centred at \vlsr=8.0 km
s$^{-1}$. No emission is detected for the \vlsr=7.0~\kms\ component, 
however 3$\sigma$ upper limit values are provided in Table \ref{tbl:overviewobsresults}.

\subsection{C$^{18}$O}
\label{sec:c18osubsection}
\begin{figure}[!t]
\centering
\includegraphics[scale=0.65]{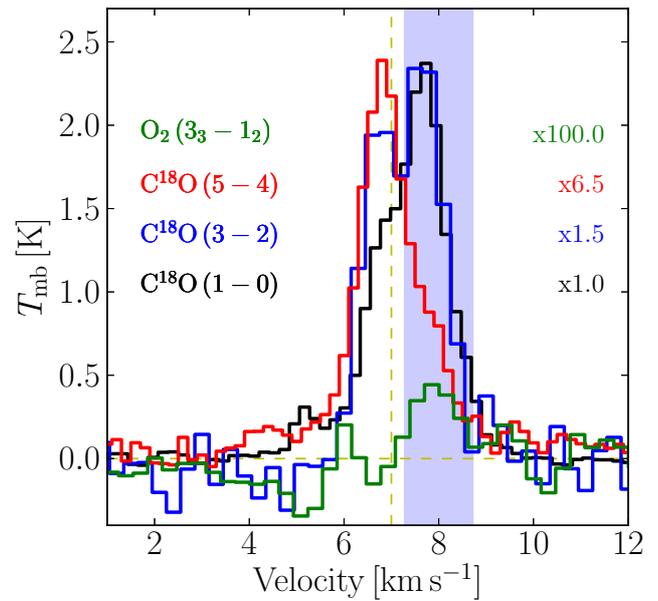}
\caption{\small{\oofull\ spectrum overplotted with the \ceio\ 1--0,
    3--2, and 5--4 lines in a 44'' beam. The \ceio\ spectra are scaled to the same
    peak intensity. Note the shift in velocity from 8.0 to 7.0~\kms\
    with increasing $J$.}}
\label{fig:isocolines}
\end{figure}

Figure~\ref{fig:isocolines} shows the \ceio\ 1--0, 3--2, and 5--4 lines
overplotted on the \oo\ line.  The peak of the
\ceio\ emission shifts from \vlsr=8.0~\kms\ to 7.0~\kms\ as $J$ increases. 
The \ceio\ 1--0 line is expected to come
primarily from the surrounding cloud at 8.0~\kms\ due to the low
energy of the transition (\Eup =~5.3~K). On the other hand, the 5--4
line has higher energy (\Eup =~79~K), therefore traces the warmer
parts of the protostellar envelope at 7.0~\kms. 
As a sanity check, the \thco\ 1--0, 3--2, and 6--5 transitions from
\citet{Yildiz12} were also inspected and their profiles are consistent
with those of the \ceio\ lines, however they are not included here due
to their high opacities. 
The integrated intensities \TmbdV\ for each of the 7.0~\kms\ and
8.0~\kms\ components are given in
Table~\ref{tbl:overviewobsresults}.

\begin{figure}[!t]
\begin{center}
\includegraphics[scale=0.25]{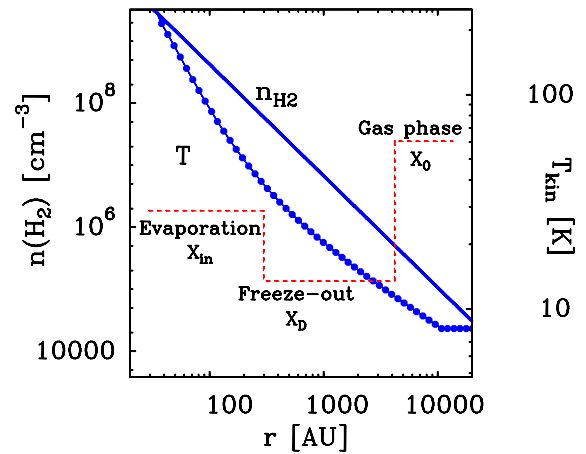}
\end{center}
\caption{\small{Variation of number density, which follows a power-law density profile and temperature of the NGC~1333 IRAS~4A envelope as
function of radial distance, taken from the model of \citet{Kristensen12}. Overplotted red dashed line shows 
the limits of drop abundance profile by radius obtained by the \ceio\ modeling as explained in Sect.~\ref{sec:abundance}.}}
\label{fig:densityTemp}
\end{figure}

\subsection{Column densities and abundances}
\subsubsection{Constant excitation temperature results}

\begin{table*}[!t]
\caption{Summary of column densities in a 44$\arcsec$ beam. See text for the conditions used for the calculations.}
\small
\begin{center}
\begin{tabular}{l c c c c c c c}
\hline \hline
Molecule  &  \multicolumn{2}{c} {Column Density [cm$^{-2}$]} &  \multicolumn{2}{c} {Abundance w.r.t.\ H$_2$} \\
 &$N$(7 \kms) & $N$(8 \kms) &$X$(7 \kms) & $X$(8 \kms) \\
\hline
\noalign{\smallskip}
O$_2$     & $<$1.2$\times$10$^{15}$\tablefootmark{(a)} & (2.8--4.3)$\times$10$^{15}$ & $\leq$5.7$\times$10$^{-9}$ & (1.3--2.1)$\times$10$^{-8}$ \\%
C$^{18}$O & (3.2--6.1)$\times$10$^{14}$              & (1.8--2.3)$\times$10$^{15}$ & (1.7-3.0)$\times$10$^{-9}$   & (4.3--2.2)$\times$10$^{-7}$ \\%
NO        & $<$1.9$\times$10$^{14}$\tablefootmark{(a)}  & 2.3$\times$10$^{14}$ & $\leq$9.0$\times$10$^{-10}$ & 2.3$\times$10$^{-8}$\\       
\noalign{\smallskip}
\hline
\noalign{\smallskip}
\hh & 2.1$\times$10$^{23}$\tablefoottext{b} & 1$\times$10$^{22}$\tablefootmark{(c)} & \dots & \dots \\%
\noalign{\smallskip}
\hline
\end{tabular}
\end{center}
\tablefoot{
\tablefoottext{a}{3$\sigma$ column density limit.}
\tablefoottext{b}{Beam averaged \hh\ column density in a 44$\arcsec$ beam obtained from the model of \citet{Kristensen12}.}
\tablefoottext{c}{Computed using the average \ceio\ column density and abundance ratios of CO/H$_2$ = 10$^{-4}$ and CO/C$^{18}$O = 550 \citep{WilsonRood94}.}
}
\label{tbl:columndensities}
\end{table*}

A first simple estimate of the \oo\ abundance limit in the IRAS~4A
protostellar envelope (\vlsr=7.0 \kms\ component) is obtained by
computing column densities within the 44$\arcsec$ beam. The
collisional rate coefficients for the \oofull\ line give a critical
density of \mbox{\ncr =1$\times$10$^{3}$~\cmthree} for low
temperatures \citep{Lique10,Goldsmith11}. 
The density at the 5000 AU radius corresponding to this beam is found
to be 4$\times$10$^{5}$~\cmthree\ based on the spherical power-law
density model of \citet[][see also Fig.~\ref{fig:densityTemp} and
below]{Kristensen12}. This value is well above the critical density,
implying that the \oo\ excitation is thermalized. High
densities are independently confirmed by the detection of many high
excitation lines from molecules with large dipole moments in this
source \citep[e.g.,][]{Jorgensen05,Maret05}.
The width of the \oofull\ line is taken to be similar to that of
C$^{18}$O, $\Delta$V$\approx$1.0 \kms.  The \oo\ line is assumed to be
optically thin and a temperature of 30~K is used.  The 3$\sigma$ \oo\
column density limit at \vlsr=7.0~\kms\ is then
$N$(\oo)=1.1$\times$10$^{15}$~\cmtwo\ assuming Equations 2
  and 3 from \citet{Yildiz12}.

The total H$_2$ column density of the 7.0 \kms\ component in the 44$\arcsec$ beam is computed from the model of
\citet{Kristensen12} through 
$N_{X,\mathrm{beam}}=\int\int n_{X}(z,b) {\rm d}z G(b) 2\pi b {\rm d}b / \int G(b)2\pi b {\rm d}b$,
where $b$ is the impact parameter, and $G(b)$ is the beam
response function. The resulting value is $N$(H$_2$)=2.1$\times$10$^{23}$ cm$^{-2}$, which is an order of magnitude lower than the
pencil-beam H$_2$ column density of 1.9$\times$10$^{24}$
cm$^{-2}$. Using the 44$\arcsec$-averaged H$_2$ column density implies an abundance
limit $X$(\oo)$\leq$5.7$\times$10$^{-9}$.  This observation therefore
provides the lowest limit on the \oo\ abundance observed to date.  It
is $\sim$4 orders of magnitude lower than the pure gas phase chemical model
predictions of \xoo\ $\sim$7$\times$10$^{-5}$.

Another option is to compare the \oo\ column density directly with
that of \ceio. These lines trace the part of the envelope where CO
and, by inference, \oo\ are not frozen out because of their similar
binding energies \citep{Collings04,Acharyya07}. Using the \ceio\ lines
therefore provides an alternative constraint on the models.  The
C$^{18}$O lines are also thermalized, and assuming a temperature of 30~K, its inferred column density
is calculated as (3.2--6.1)$\times$10$^{14}$ \cmtwo, depending on the adopted lines.
The corresponding abundance ratio is $N$(\oo)/$N$(\ceio)=$\leq$3.5
so $N$(\oo)/$N$(\co)$\leq 6.4 \times$10$^{-3}$ assuming
CO/C$^{18}$O=550.  

The critical densities for the NO transitions 
range from $n_{\rm cr\,NO(1)}$=2.4$\times$10$^{4}$
cm$^{-3}$ to $n_{\rm cr\,NO(4)}$=7.0$\times$10$^{3}$ cm$^{-3}$),
so LTE is again justified. For the 3$\sigma$
upper limit on the NO~(3) line in the 7.0~\kms\ component, the
inferred column density is
$N$(NO)=$<$1.9$\times$10$^{14}$~\cmtwo, assuming $T_{\rm kin}$=30~K and no
beam dilution. 
Thus, the implied NO abundance is
$N$(NO)/$N$(\hh)=$X$(NO)$\leq$9.0$\times$10$^{-10}$.
All column densities and abundances associated with the protostellar
source at \vlsr~=~7.0~\kms\ are summarized in
Table~\ref{tbl:columndensities}.

\subsubsection{Abundance variation models}
\label{sec:abundance}

The above analysis assumes constant physical conditions along the line
of sight as well as constant abundances. It is well known from
multi-line observations of C$^{18}$O that the CO abundance varies
throughout the envelope, dropping by more than an order of magnitude
in the cold freeze-out zone
\citep[e.g.,][]{Jorgensen02,Yildiz10,Yildiz12}.  A more sophisticated
analysis of the \oo\ abundance is therefore obtained by using a model
of the IRAS~4A envelope in which the density and temperature vary with
position. The envelope structure presented in
Fig.~\ref{fig:densityTemp} has been determined by modeling the
continuum emission (both the spectral energy distribution and the
submillimeter spatial extent) using the 1D spherically symmetric dust
radiative transfer code \verb1DUSTY1 \citep{IvezicElitzur97}.  A
power-law density profile is assumed with an index $p$, i.e., $n
\propto r^{-p}$, and the fitting method is described in
\citet{Schoier02} and \citet{Jorgensen02,Jorgensen05}, and is further
discussed in \citet{Kristensen12} with the caveats explained.
The temperature is calculated as a function of position by solving for
the dust radiative transfer through the assumed spherical envelopes,
heated internally by the luminosity of the source.  The gas
temperature is assumed to be equal to the dust temperature.
The envelope is defined from the inner radius of 33.5~AU up to
the outer radius of 33\,000~AU, where the density of the outer
radius is 1$\times$10$^{4}$~\cmthree. IRAS~4A is taken to be a
standalone source; the possible overlap with IRAS~4B is ignored,
but any material at large radii along the line of sight within
the beam contributes in both the simulated and observed spectra.

\begin{figure*}[!t]
\centering
\includegraphics[scale=0.51]{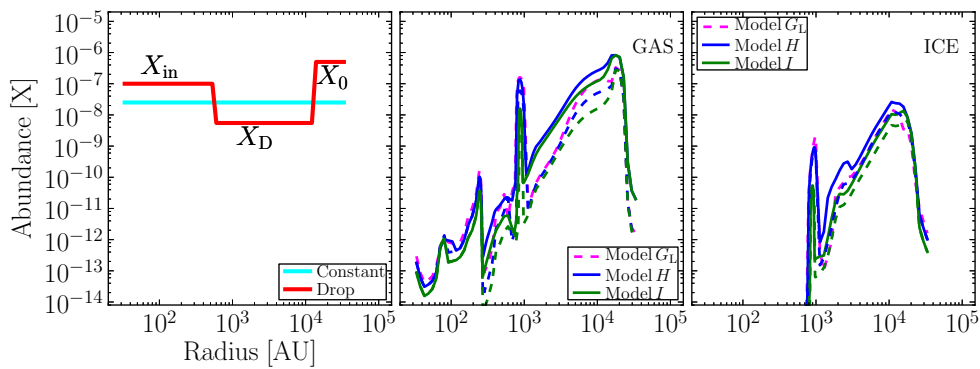}
\caption{\small{{\it Left:} Schematic diagram showing the best-fit drop abundance
    profile for \oo\ (red line) assuming \oo\ follows the same freeze-out and sublimation processes as \ceio. A constant (light blue) 
abundance profile is also shown.
    {\it Middle:} best-fit gaseous \oo\ profiles obtained via gas-grain modeling. Models $H_{\rm O}$ and $I_{\rm O}$ are shown in solid lines and models $H_{\rm L}$, $I_{\rm L}$ and $G_{\rm L}$ using a lower O + OH rate coefficient 
in dashed lines. As shown
in Fig.~\ref{fig:iras4aO2models} these are the maximum abundances of gaseous \oo\ that can be hidden in the spectrum.
{\it Right:} \oo\ ice abundances for the best fit models. Solid lines show $H_{\rm O}$ and $I_{\rm O}$ whereas dashed lines 
are Models $G_{\rm L}$, $H_{\rm L}$ and $I_{\rm L}$. }}
\label{fig:iras4aprofiles}
\end{figure*}

\begin{figure}[!t]
\centering
\includegraphics[scale=0.32]{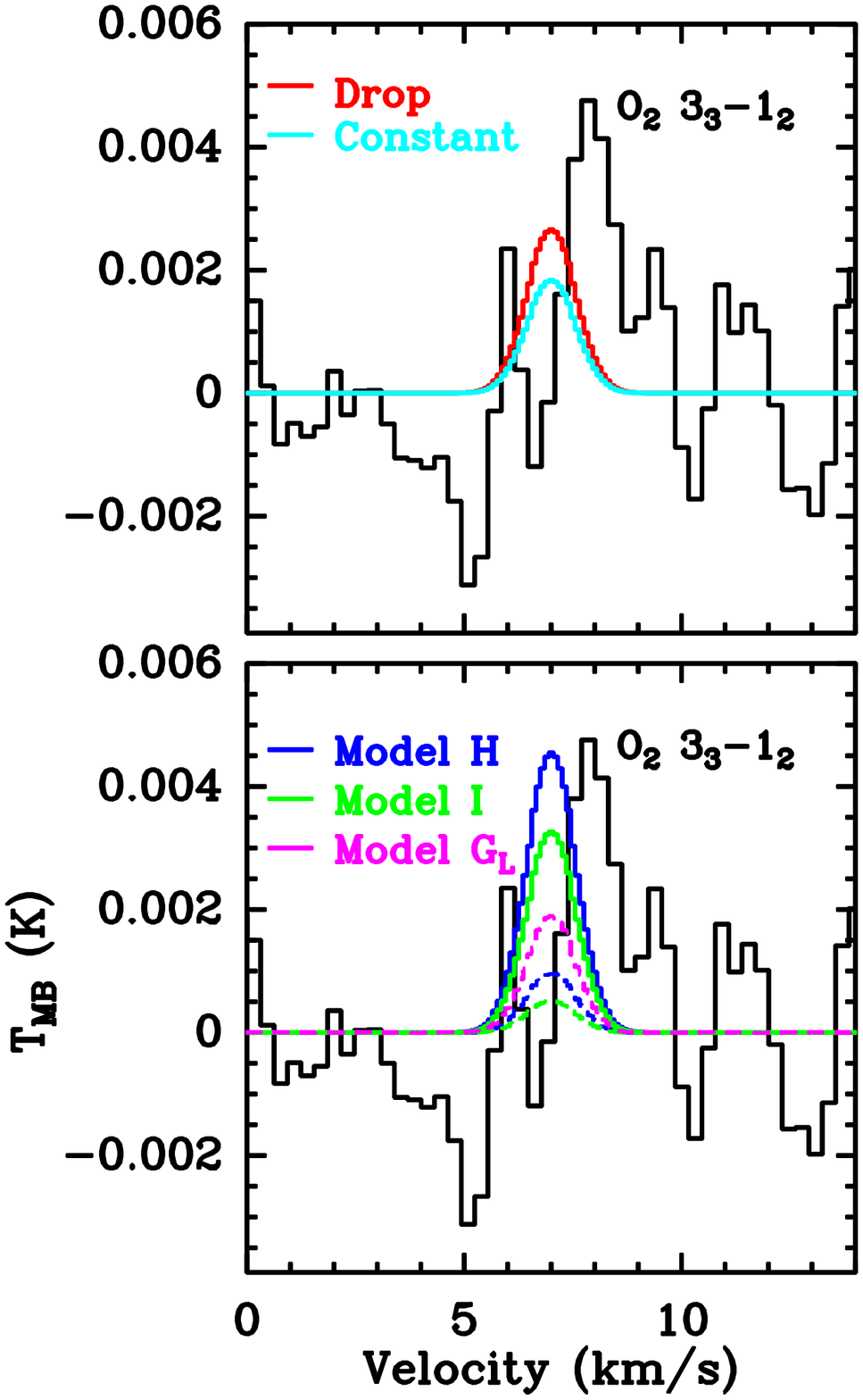}
\includegraphics[scale=0.32]{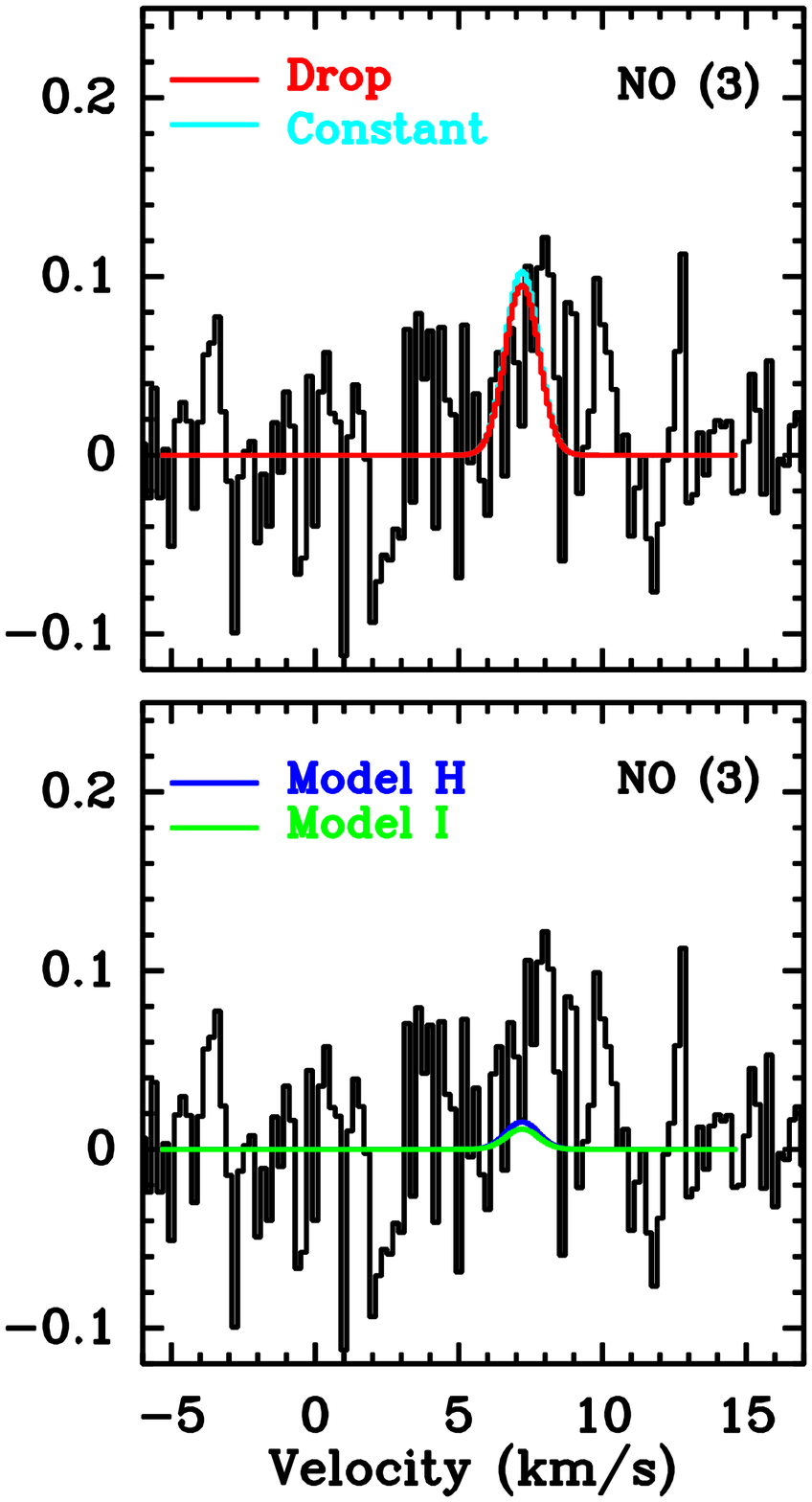}
\caption{\small{Best-fit model spectra produced by different abundance 
profiles for \oofull\ and NO\,(3) are overplotted over the observed spectra. 
In the buttom figures, solid fit spectra show $H_{\rm O}$ and $I_{\rm O}$ 
models and dashed fit spectra show models $H_{\rm L}$, $I_{\rm L}$, and 
$G_{\rm L}$}. For NO, Case 2 is adopted.}
\label{fig:iras4aO2models}
\end{figure}

The observed line intensities are used to constrain the molecular
abundances in the envelope by assuming a trial abundance structure and
computing the non-LTE excitation and line intensities with radiative
transfer models for the given envelope structure. For this purpose,
the Monte Carlo line radiative transfer program \verb1Ratran1
\citep{Hogerheijde00} is employed.  The simplest approach assumes a
constant \oo\ abundance through the envelope.
Figure~\ref{fig:iras4aprofiles} ({\it left}) shows different abundance
profiles, whereas Fig.~\ref{fig:iras4aO2models} ({\it top left}) shows
the resulting line intensities overplotted on the observed \oo\
line. The light blue line in Figs.~\ref{fig:iras4aprofiles} 
and \ref{fig:iras4aO2models} is the maximum constant \oo\ abundance
that can be hidden in the noise, which is 2.5$\times$10$^{-8}$. This is
within a factor of 4 of the simple column density ratio estimate.

A more realistic abundance structure includes a freeze-out zone below
25 K where both O$_2$ and CO are removed from the gas. Such a CO
`drop' abundance profile has been determined for the IRAS~4A envelope
via the optically thin \ceio\ lines from 1--0 to 10--9 in
\citet{Yildiz12}. By using the best fit CO abundance structure and
assuming a constant O$_2$/CO abundance ratio, an upper limit of 
\oo/\ceio$\leq$1 is obtained (see red line in Figs.~\ref{fig:iras4aprofiles} 
and \ref{fig:iras4aO2models}), corresponding to 
O$_2$/CO$\leq$2$\times$10$^{-3}$.

With a 44$\arcsec$ beam, the 487~GHz line observed with HIFI is mostly
sensitive to the bulk of the envelope. Nevertheless, the drop
abundance models can be used to estimate the maximum \oo\ abundance on
smaller scales, to get a firm observational constraint on how much
\oo\ is in the region where it could enter the embedded circumstellar
disk. The radius of such a disk is highly uncertain, but probably on
the order of 100~AU \citep[e.g.][]{Visser09}. According to the drop
abundance models, the maximum \oo\ abundance that can be ``hidden''
inside 100~AU is $\sim$10$^{-6}$. However, the full chemical models
from Sect.~\ref{sec:modeling} suggest the actual \oo\ abundance on
these small scales is several orders of magnitude lower
(Fig.~\ref{fig:iras4aprofiles}, middle).

In summary, both the simple column density estimate and the more
sophisticated envelope models imply a maximum \oo\ abundance of
$\sim$10$^{-8}$, and an O$_2$/CO ratio of $\leq$2$\times$10$^{-3}$.
For NO, the best fit drop abundance requires NO to be about 8
times lower in abundance than \ceio, to be consistent with our NO
non-detection.

\section{Gas-grain models for the protostellar envelope}
\label{sec:modeling}

The next step in the analysis is to compare the upper limit for the
\vlsr=7.0 km s$^{-1}$ component with full gas-grain chemical models. The
Ohio State University (OSU) gas-grain network \citep{Garrod08} is used as the
basis for the chemical network, which contains an extensive gas-grain
chemistry. There are 590 gas phase and 247 grain surface species and
7500 reactions among them. 
The subsequent subsections discuss the various chemical processes
chemical processes that were considered in the network and are
relevant for \oo\ and NO are discussed.

\subsection{Gas phase \oo\ and \no\ formation}

In the gas, \oo\ is predominantly formed via 
reaction (\ref{eq:o2reaction}) between O atoms and OH radicals. 
The rate coefficient of this reaction has been measured in the
temperature range between 39~K and 142~K with the CRESU (Cinetique de
Reaction en Ecoulement Supersonique Uniforme) technique by
\cite{Carty06} who found a rate coefficient of 3.5$\times$10$^{-11}$
cm$^3$ s$^{-1}$ that is constant with temperature. However, several
theoretical calculations, especially below 39~K, also exist in the
literature.  Using quantum mechanical calculations with the so-called
$J$-shifting approximation and neglecting non-adiabatic coupling,
\cite{Xu07} obtained a rate coefficient that decreases as the
temperature drops from 100 to 10~K. At 10~K, the computed rate
coefficient has fallen to a value of 5.4$\times$10$^{-13}$
cm$^3$~s$^{-1}$, significantly lower than the 39 K experimental
value. However, more recent calculations by \citet{Lin08}, in which
the $J$-shifting approximation has been removed, find a rate
coefficient at 10~K of 7.8$\times$10$^{-12}$ cm$^3$~s$^{-1}$, higher
than the \citet{Xu07} value but still only about 1/4.5 of the
experimental value at 39~K.  The \oo\ formation rates are summarised
in Table~\ref{tbl:rates}.  We have used all three rate coefficients
to compare the computed \oo\ abundance with the measured
upper limit.

Gas-phase NO is predominantly formed 
through reaction (\ref{eq:noreaction}).
Its gas-phase reaction rate coefficient is listed in Table~\ref{tbl:rates}. This
reaction is taken from the OSU database and was first determined by
\citet{Smith04}.

\begin{table}[!t]
\caption{Rate coefficients for \oo\ (Equation \ref{eq:o2reaction}) and NO (Equation \ref{eq:noreaction}) formation.}
\tiny
\begin{center}
\begin{tabular}{l r r r l l}
\hline \hline
No. & Species & $T$~[K] & Rate coeff.\ [cm$^3$ s$^{-1}$] & References  \\
\hline
\noalign{\smallskip}
1.\tablefootmark{a} & O$_2$  & 39--149 & 3.5$\times$10$^{-11}$                  & \citet{Carty06} \\
2.\tablefootmark{b} & O$_2$  & 10      & 7.8$\times$10$^{-12}$                  & \citet{Lin08}  \\
3.\tablefootmark{c} & O$_2$  & 10      & 5.4$\times$10$^{-13}$                  & \citet{Xu07}  \\
4.                  & O$_2$  & \dots   & 7.5$\times$10$^{-11}$$\times$(T/300)$^{-0.25}$ & OSU database \\
5.                  & NO     & \dots   & 7.5$\times$10$^{-11}$$\times$(T/300)$^{-0.18}$ & OSU database  \\
\noalign{\smallskip}
\hline
\end{tabular}
\end{center}
\tablefoot{
\tablefoottext{a}{CRESU measurement}
\tablefoottext{b}{without $J$-shifting}
\tablefoottext{c}{with $J$-shifting}
}
\label{tbl:rates}
\end{table}

\begin{table*}[!t]
\centering
\caption{Chemical models considered for the IRAS 4A protostellar envelope}
\begin{tabular}{l r r r r r r r}
\hline \hline
Model & Pre-collapse & Protostellar  & O$_2$ formation rate  & \Tpeak (\oo)\tablefootmark{a} & \Tpeak (\no)\tablefootmark{a}   \\
      & stage [yr]       & stage [yr]         & [cm$^3$~s$^{-1}$]           & [K]  & [K] \\
\hline
\noalign{\smallskip}
$A$  & 5$\times$10$^{4}$  & 10$^5$   & 7.5$\times$10$^{-11}$$\times$(T/300)$^{-0.25}$ & 0.119\phantom{0}  & \dots \\
$B$  & 1$\times$10$^{5}$  & 10$^5$   & 7.5$\times$10$^{-11}$$\times$(T/300)$^{-0.25}$ & 0.102\phantom{1}   & \dots \\
$C$  & 2$\times$10$^{5}$  & 10$^5$   & 7.5$\times$10$^{-11}$$\times$(T/300)$^{-0.25}$ & 0.085\phantom{1}     & \dots \\
$D$  & 3$\times$10$^{5}$  & 10$^5$   & 7.5$\times$10$^{-11}$$\times$(T/300)$^{-0.25}$ & 0.073\phantom{1}    & \dots \\
$E$  & 5$\times$10$^{5}$  & 10$^5$   & 7.5$\times$10$^{-11}$$\times$(T/300)$^{-0.25}$ & 0.042\phantom{1}    & \dots \\
$F$  & 6$\times$10$^{5}$  & 10$^5$   & 7.5$\times$10$^{-11}$$\times$(T/300)$^{-0.25}$ & 0.024\phantom{0}    & \dots \\
$G_{\rm O}$  & 7$\times$10$^{5}$  & 10$^5$   & 7.5$\times$10$^{-11}$$\times$(T/300)$^{-0.25}$ & 0.011\phantom{1}    & \dots \\
$G_{\rm L}$\tablefootmark{b}  & 7$\times$10$^{5}$  & 10$^5$   & 7.84$\times$10$^{-12}$ &  0.0019 & \dots   \\
$H_{\rm O}$\tablefootmark{b}  & 8$\times$10$^{5}$  & 10$^5$   & 7.5$\times$10$^{-11}$$\times$(T/300)$^{-0.25}$ & 0.0046    & 0.015 \\
$H_{\rm C}$\tablefootmark{b}  & 8$\times$10$^{5}$  & 10$^5$   & 3.50 $\times$10$^{-11}$ & 0.0025 & \dots  \\
$H_{\rm L}$\tablefootmark{b}  & 8$\times$10$^{5}$  & 10$^5$   & 7.84 $\times$10$^{-12}$ & 0.0010 & \dots   \\
$I_{\rm O}$\tablefootmark{b}  & 1$\times$10$^{6}$  & 10$^5$   & 7.5  $\times$10$^{-11}$$\times$(T/300)$^{-0.25}$ &  0.0033 & 0.011 \\
$I_{\rm C}$\tablefootmark{b}  & 1$\times$10$^{6}$  & 10$^5$   & 3.50 $\times$10$^{-11}$ & 0.0014 & \dots  \\
$I_{\rm L}$\tablefootmark{b}  & 1$\times$10$^{6}$  & 10$^5$   & 7.84 $\times$10$^{-12}$ & 0.0005 & \dots   \\
$J$  & 8$\times$10$^{4}$  & 2$\times$10$^4$ & 3.50 $\times$10$^{-11}$ &  0.106\phantom{1}   & \dots \\
$L$  & 3$\times$10$^{5}$  & 10$^5$   & 3.50$\times$10$^{-11}$ & 0.063\phantom{1}    & \dots \\
$M$  & 3$\times$10$^{5}$  & 10$^5$   & 7.84$\times$10$^{-12}$ & 0.047\phantom{1}    & \dots \\
$N$  & 3$\times$10$^{5}$  & 10$^5$   & 5.40$\times$10$^{-13}$ & 0.018\phantom{1}    & \dots \\
$O$  & 5$\times$10$^{5}$  & 10$^5$   & 3.50$\times$10$^{-11}$ & 0.031\phantom{1}    & \dots \\
$P$  & 5$\times$10$^{5}$  & 10$^5$   & 7.84$\times$10$^{-12}$ & 0.020\phantom{1}    & \dots \\
$Q$  & 5$\times$10$^{5}$  & 10$^5$   & 5.40$\times$10$^{-13}$ & 0.0068   &  0.091 \\
\noalign{\smallskip}
\hline
\end{tabular}
\tablefoot{
{Models $G$, $H$, and $I$ use different rate coefficients for \oo\ formation (Eq.~\ref{eq:o2reaction}), as indicated by subscripts: O for OSU, C for \citet{Carty06}, and L for \citet{Lin08}.}
\tablefoottext{a}{Ratran model results using a line width of 1.0 km s$^{-1}$.}
\tablefoottext{b}{Best fit models. For NO, Case 2 is tabulated.}
}
\label{tbl:modelsAQ}
\end{table*}

\subsection{Grain chemistry specific to \oo\ and NO}
The grain surface chemistry formulation in the OSU code follows the
general description by \citet{Hasegawa93} for adsorption, diffusion,
reaction, dissociation, and desorption processes, updated and extended
by \citet{Garrod08}. 
The binding energies of various species to the surface are critical
parameters in the model.  In most of the models, we adopt the binding
energies from \citet{Garrod06} appropriate for a water-rich ice
surface. However, the possibility of a CO-rich ice surface is also
investigated by reducing the binding energies by factors of 0.75 and
0.5, respectively \citep[][]{Bergin95,Bergin97}.

The presence of \oo\ on an interstellar grain can be attributed
to two different processes. First, gas-phase \oo\ can be accreted on
the grain surface during the (pre-)collapse phase and second, atomic
oxygen can recombine to form \oo\ on the dust grain via the following
reaction:
\begin{equation}
{\rm O + O \rightarrow O_2}.
\end{equation}
Following \citet{Tielens82}, 800~K is used as the binding
(desorption) energy for atomic oxygen on water ice. 
The binding energy for \oo\ on water ice is taken as 1000~K 
\citep{Cuppen07}, which is an average value obtained from
the temperature programmed desorption (TPD) data by \citet{Ayotte01} and
\citet{Collings04}. A ratio of 0.5 between the diffusion 
barrier and desorption energy has been
assumed for the entire calculation \citep{Cuppen07}, so the 
hopping energy for atomic oxygen is 400~K.

For this hopping energy, one oxygen atom requires 2$\times$10$^5$
seconds to hop to another site at 10~K.  For comparison, the time needed for a
hydrogen atom to hop to another site is around 0.35 seconds, which is
a factor of 10$^6$ faster. Therefore, instead of forming \oo,
an accreted atomic oxygen species will be hydrogenated, 
leading to the formation of OH and \hho.
It is most unlikely that accreted atomic oxygen produces
any significant amount of \oo\ on the grain surface during the
pre-collapse phase. Recent
studies using the continuous time random walk (CTRW) Monte Carlo
method do not produce significant \oo\ on the grain surface \citep{Cuppen07}.
However, elevated grain temperatures ($\gtrsim$20~K), when the residence 
time of an H atom on the grain is very short and atomic oxygen has enhanced 
mobility, could be conducive to \oo\ formation.

What happens to the \oo\ that is formed in the gas phase and accreted
onto the dust grains?  There are two major destruction pathways.
First, the reaction of O$_2$ with atomic H leads to the formation of
HO$_2$ and H$_2$O$_2$, which then could be converted to water
following reaction pathways suggested by \citet{Ioppolo08} and
\citet{Cuppen10}:
\begin{equation} 
\rm {O_2 \buildrel H \over \longrightarrow HO_2 \buildrel H \over \longrightarrow H_2O_2 \buildrel H \over \longrightarrow H_2O + OH.} 
\end{equation}
Thus, a longer cold pre-collapse phase would significantly reduce
\oo\ on the dust grains and turn it into water ice, whereas a shorter
pre-collapse phase would yield a higher solid \oo\ abundance \citep{Roberts02}.
These reactions also depend on the grain temperature: at higher temperatures, 
the shorter residence time of H atoms on the grain leads to less conversion of \oo.

The second destruction route leads to the formation of ozone through
\begin{equation}
{\rm O_2 + O \rightarrow O_3.}
\end{equation}
This route is most effective at slightly higher grain temperatures ($\gtrsim$20~K)
when atomic oxygen has sufficient mobility to find an \oo\ molecule before
it gets hydrogenated.
 Ozone could also be hydrogenated as suggested by
\citet{Tielens82} and confirmed in the laboratory by
\citet{Mokrane09} and \citet{Romanzin10} leading back to \oo:
\begin{equation} 
\rm {O_3 \buildrel H \over \longrightarrow O_2 + OH.}
\end{equation}

Similarly, accreted NO on the grain surface can undergo various
reactions. In particular, recent laboratory experiments have shown
that NO is rapidly hydrogenated to NH$_2$OH at low ice temperatures
\citep{Congiu12}. A critical parameter here is the competition of the
different channels for reaction of HNO + H, which can either go back
to NO + H$_2$ or form H$_2$NO.

The final important ingredient of the gas-grain chemistry is the rate
at which molecules are returned from the ice back into the gas phase.
Both thermal and non-thermal desorption processes are considered.  The
first non-thermal process is reactive desorption; here the exothermicity 
of the reaction is channeled into the desorption of the product with an 
efficiency determined by a parameter $a_{\rm RRK}$ \citep{Rob07}. In 
these model runs, a value of 0.01 is used, which roughly translates into 
an efficiency of 1\%. Recently, \citet{Du12} used 7\% for the formation
of H$_2$O$_2$.  

Second, there is desorption initiated by UV absorption.
Photodissociation of an ice molecule produces two atomic or radical
products, which can subsequently recombine and desorb via the
reactive desorption mechanism. The photons for this process derive
both from the external radiation field and from UV photons generated
by ionization of \hh\ due to cosmic rays, followed by the excitation of
\hh\ by secondary electrons. The externally generated
UV photons are very effective in diffuse and translucent
clouds but their role in dense clouds is limited to the edge of the
core \citep{Ruf00,Hollenbach09}. The cosmic-ray-generated internal photons
can play an effective role in the dense envelope, with a photon flux
of $\approx$10$^4$ photons cm$^{-2}$~s$^{-1}$ \citep{Shen04}. We
have considered both sources of radiation in our model. In either
case, the rate coefficients for photodissociation on surfaces are
assumed to be the same as in the gas phase. 

Photodesorption can proceed both by the recombination mechanism
described above as well as by kick-out of a neighboring molecule. The
combined yields for a variety of species including CO and H$_2$O have been
measured in the laboratory \citep{Obe09b,Obe09a,Munoz10} and computed
through molecular dynamics simulations for the case of H$_2$O by
\citet{Andersson08} and \citet{Arasa10}. Finally, there is the heating of
grains via direct cosmic ray bombardment, which is effective for
weakly bound species like CO and O$_2$ and included following the
formulae and parameters of \citet{Hasegawa93}.

\subsection{Model results}
\label{sec:modelresults}

Our physical models have two stages, the ``pre-collapse stage'' and the
``protostellar stage''.  In the pre-collapse stage, the hydrogen
density is $n_{\rm H}$=10$^5$ \cmthree, visual extinction $A_{\rm
  V}$=10 mag, the cosmic-ray ionization parameter, $\zeta$=
1.3$\times$10$^{17}$~s$^{-1}$, and the (gas and grain) temperature,
$T$=10~K, which are standard parameters representative of cold cores.  The
initial elemental abundances of carbon, oxygen and nitrogen are
7.30$\times$10$^{-5}$, 1.76$\times$10$^{-4}$ and
2.14$\times$10$^{-5}$, respectively, in the form of atomic C$^+$, O
and N.  All hydrogen is assumed to be in molecular form initially.  In
the second stage, the output abundance of the first phase is used as
the initial abundance at each radial distance with the density,
temperature and visual extinction parameters at each radius taken from
the IRAS~4A model shown in Fig.~\ref{fig:densityTemp}. We
assume that the transition to the protostellar phase from the
pre-collapse stage is instantaneous i.e., the power-law density and
temperature structure are established quickly, consistent with
evolutionary models \citep{Lee04,Young05}.

To explain the observed spectra of \oo, both the
pre-collapse time and protostellar time as well as the \oo\ formation
rates are varied. Analysis of CO and HCO$^+$ multi-line observations
in pre- and protostellar sources have shown that the high density
pre-collapse stage typically lasts a few $\times$10$^5$ yr
\citep[e.g.,][]{Jorgensen05,Wardthompson07}. The models $A$ to $Q$ have
different parameters and timescales which are listed in
Table~\ref{tbl:modelsAQ}. Those models result in abundance profiles in
the envelope at each time step and radius. These profiles are then run
in \verb1Ratran1 in order to compare directly with the observations.

Figure~\ref{fig:iras4aprofiles} ({\it middle}) shows examples of model
abundance profiles. 
All model runs predict lower \oo\ abundances than the evolutionary models of 
\citet{Visser11}, whose chemical network did not include any grain-surface 
processing of \oo. The line emission from our models is compared 
to the observations in Fig. 8 (bottom left).
Table~\ref{tbl:modelsAQ} summarizes the resulting \oo\ peak
temperatures for each of the models.  All models except $H$ and $I$ 
overproduce the observed \oo\ emission of at most a few mK, by up
to two orders of magnitude in the peak temperature.  The models that are
consistent with the data have in common longer pre-collapse stages; in
particular Model $I$, which has the longest pre-collapse stage of
10$^{6}$~years, best fits the 3$\sigma$ \oo\ limit. Using a lower rate 
coefficient for the 
O+OH reaction of 7.84$\times$10$^{-12}$ \mbox{cm$^3$~s$^{-1}$} \citep{Lin08} 
for the same best fit models ($H_{\rm L}$ and $I_{\rm L}$) 
results in a factor of 5 lower emission. In this case, the pre-collapse 
stage can be shortened to $\sim$7$\times$10$^{5}$~yr (Model $G_{\rm L}$).

For NO, the $H$ and $I$ models were calculated twice. In Case~1,
hydrogenation of HNO leads back to NO and \hh\ and in Case~2
hydrogenation of HNO leads to H$_2$NO as suggested by
\citet{Congiu12}. Comparison of the results from Case 1 with
observations shows significant overproduction of the observed NO
emission, whereas Case~2 is consistent with an upper limit of 0.05~K (1~$\sigma$). Therefore, a
combination of both reactions appears to be needed.

These results for IRAS~4A suggest that a long pre-collapse stage is
characteristic of the earliest stages of star formation, in
which atomic and molecular oxygen are frozen-out onto the dust grains
and converted into water ice, as proposed by \citet{Bergin00}.
Similarly, the rapid conversion of NO to other species on the grains
limits its gas-phase abundance. It is clear that the 
grain surface processes are much more important than those of the pure
gas-phase chemistry in explaining the \oo\ and NO observations. The timescale
for NGC~1333 IRAS~4A is at the long end of that inferred more generally from observations.

The model results show that the fraction of \oo\ in the gas and left on the
grains must indeed be very small, dropping to $\leq$10$^{-9}$ close to
the protostar (Fig.~\ref{fig:iras4aprofiles}, {\it right}). This in turn
implies that the gas and ice that enter the disk are very poor in \oo.
Although IRAS~4A is the only low-mass protostar that has been observed to this 
depth, the conclusions drawn for IRAS~4A probably hold more generally. Thus, 
unless there is significant production of \oo\ in the disk, the icy planetesimals 
will also be poor in \oo.

\section{Tentative detection of \oo\ in the 8~\kms\ cloud}
\label{sec:tentative}

\begin{figure}[!t]
\begin{center}
\includegraphics[scale=0.33]{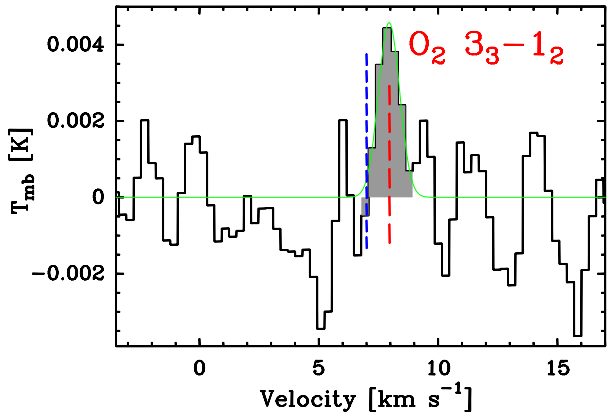}
\end{center}
\caption{\small{Tentative detection of the \oofull\ line from the
    extended NGC~1333 cloud. The blue line indicates the source
    velocity of IRAS~4A at 7.0~\kms\ and the red line indicates the
    extended cloud \vlsr\ at 8.0~\kms. The green line indicates a Gaussian fit to the component at 8.0~\kms.}}
\label{fig:tentative}
\end{figure}

\begin{figure}[!t]
\centering
\includegraphics[scale=0.45]{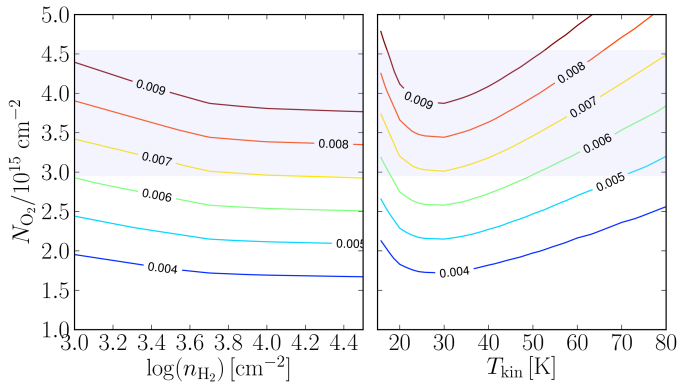}
\caption{\small{Integrated intensities of O$_2$ are shown as
      contours for a range of  model column densities as function of density ({\it
        left}) and temperature ({\it right})  for the 8.0 \kms\
      component. The yellow solid line is the observed integrated
      intensity of the tentative detection and the shaded region
      represents the range of column densities for their
      corresponding temperatures. {\it Left:} $T_{\rm kin}$=30~K is
      assumed for different densities and {\it right:}
      $n$=5$\times$10$^{3}$~cm$^{-3}$ is assumed for different
      temperatures.}}
\label{fig:intintcontours}
\end{figure}

Although there is no sign of \oo\ emission at the velocity of the
dense protostellar envelope (7.0~\kms), a 4.5$\sigma$ tentative
detection of \oofull\ line emission is found at \vlsr=8.0~\kms, the
velocity of the more extended NGC 1333 molecular cloud
\citep[Fig.~\ref{fig:tentative};][]{Loren76,Liseau88}. 
The feature is also seen in individual  H and V polarization spectra.
The peak intensity of the tentative detection is $T_{\rm mb}$=4.6 mK, 
the line width $\Delta V$=1.3~\kms, and the integrated intensity is $\int T_{\rm mb} 
{\rm d}V$=6.9 mK~km~s$^{-1}$ between the velocities of 6.5 to 9.7~\kms.
The large HIFI beam size of 44$\arcsec$ encompasses both the extended
cloud as well as the compact envelope. Since any \oo\ emission is
optically thin, the two components cannot block each other, even if
slightly overlapping in velocity.

The density in the surrounding cloud at 8.0~\kms\ is expected to be
significantly lower than that in the envelope. Figure~\ref{fig:radex}
(in the Appendix) presents the \ceio\ 3--2/1--0 ratio for the
8.0~\kms\ component. The observed value is consistent with a wide
range of kinetic temperatures from 20~K to 70~K, with corresponding
densities in a narrow range from 7$\times$10$^3$ to 2$\times$10$^3$
\cmthree, respectively.  \ceio\ column densities from the \mbox{3--2} and
\mbox{1--0} line intensities are then (2.3--1.8)$\times$10$^{15}$ \cmtwo\ for
this range of physical parameters.

For the same range of conditions, the \oo\ column density is
(2.8--4.3)$\times$10$^{15}$~\cmtwo\ 
(Fig.~\ref{fig:intintcontours}). The inferred abundance
ratios are $N$(\oo)/$N$(\ceio)=1.2 to 2.4, and
$N$(\oo)/$N$(\co)=(2.2--4.3)$\times$10$^{-3}$. Assuming
CO/H$_2$$\approx$10$^{-4}$ gives
$N$(\oo)/$N$(\hh)=(2.2--4.3)$\times$10$^{-7}$.  Interestingly, the
inferred \oo\ abundance is in between the values found for the Orion
and $\rho$ Oph A clouds \citep{Goldsmith11,Liseau2012}.  
By assuming $T_{\rm kin}$=30~K, the implied NO column density
is 2.3$\times$10$^{14}$~\cmtwo, leading to
$N$(NO)/$N$(\oo)=(5.3--8.1)$\times$10$^{-2}$ and 
$N$(NO)/$N$(\hh)=2.3$\times$10$^{-8}$.

\begin{figure}[!t]
\centering
\includegraphics[scale=0.6]{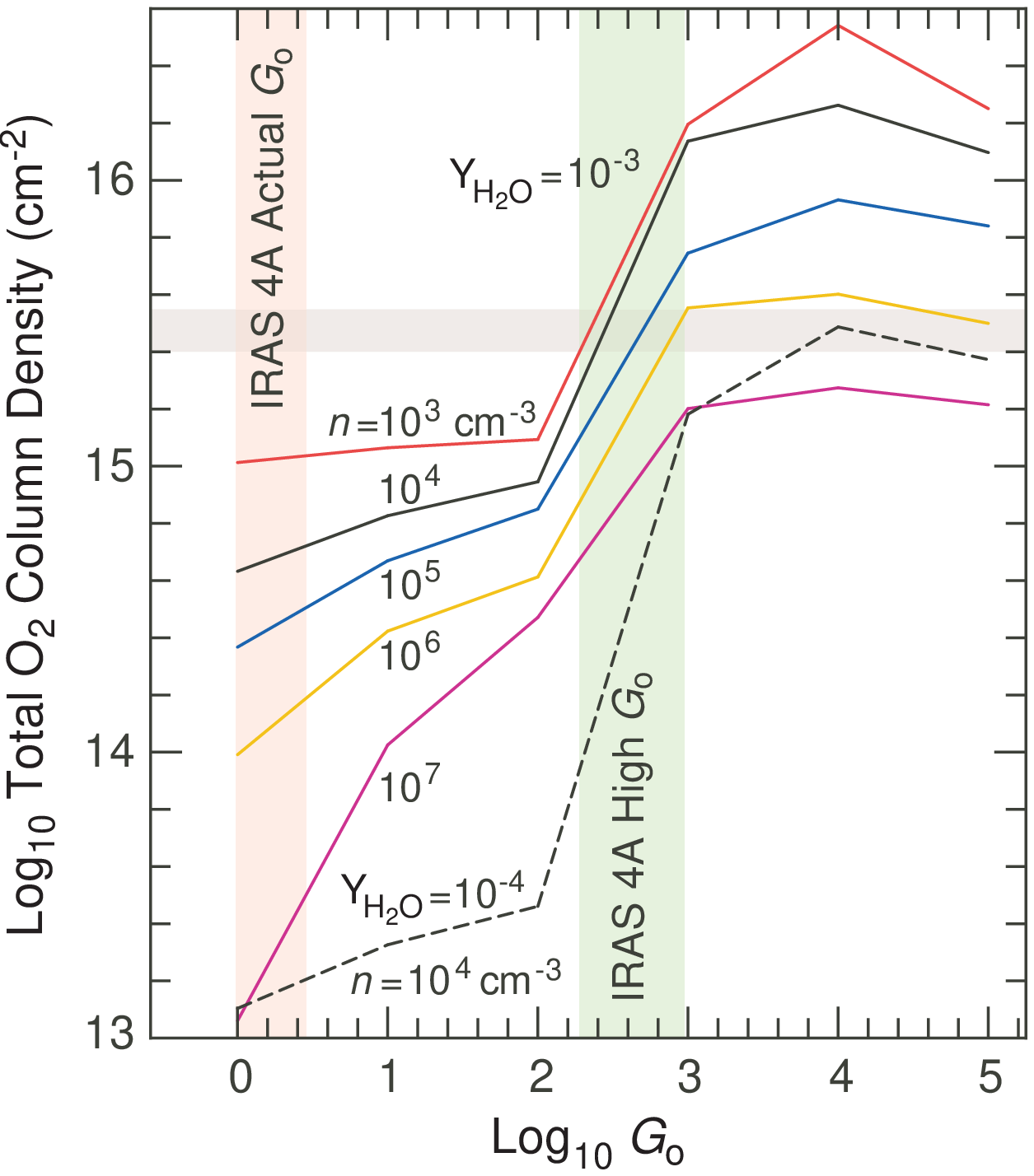}
\caption{\small{Total \oo\ column density as a function of $G_o$ and
density \citep[adapted from][]{Melnick12, Hollenbach09}
for the 8.0 \kms\ component. The horizontal grey band
shows the total \oo\ column density range for the observed
integrated intensity in the 8 km s$^{-1}$ component. The vertical
green band presents the range of high-\G0\ values required to
produce this range of \oo\ column densities for gas densities
between 10$^{3}$~\cmthree\ and 10$^{7}$~\cmthree, whereas the orange
band presents the actual \G0. $Y_{\rm H_{2}O}$ is the water ice
photodesorption yield.}}
\label{fig:iras4aG0}
\end{figure}

Can such an \oo\ column density and abundance be reproduced by
chemical models? For the surrounding cloud, a large gas-grain model is
not needed. Instead, the PDR models of \citet{Hollenbach09}, which
include a simplified gas-grain chemistry, are adequate to model the
emission. Figure~\ref{fig:iras4aG0} is a plot adapted from
\citet{Melnick12}, which shows the values of interstellar radiation field \G0\ required to
reproduce the range of \oo\ column densities according to the model
described in \citet{Hollenbach09}. The horizontal grey band bounds the
total observed \oo\ column density range of
(2.8--4.3)$\times$10$^{15}$~\cmtwo, while the vertical green band shows
the range of \G0\ values required to produce this range of \oo\ column
densities for gas densities between 10$^{3}$~\cmthree\ and
10$^{7}$~\cmthree. For our low inferred densities of <10$^4$ \cmthree, a high \G0\ value
of 300--650 fits the data.  There is no external source in the
NGC~1333 region which can provide this level of UV illumination, not
even the nearby B8~V type star BD+30$\degr$549
\citep[$\alpha$=$3^{\mathrm{h}}29^{\mathrm{m}}19\fs78$,
$\delta$=$+31\degr24\arcmin57\farcs05$ (J2000); $\sim$0.8~pc
away;][]{vanLeeuwen07}. The \G0\ value from this star at our position
is at most 2.8 and can therefore not explain the high-\G0\ scenario.
Models with a slightly higher (factor of 2) water ice photodesorption
yield than the standard value of $Y_{\rm H_{2}O}$=10$^{-3}$ would fit
better the low-\G0\ regime, where the value of \G0\ is between the
interstellar value of 1 and 2.8.  This value of $Y_{\rm H_{2}O}$ 
is within the uncertainties of the laboratory
\citep{Obe09b} and theoretical work \citep{Arasa10}. More generally,
a factor of two uncertainty in abundance can readily result from the
combined uncertainties of the individual rate coefficients that lead
to O$_2$ formation and destruction \citep{Wakelam06}, so it does not
necessarily imply an increased value of $Y_{\rm H_{2}O}$. 
Considering the uncertainties in rate coefficients as well as possible 
increase in $Y$, there does not seem to be a problem having the actual low 
value of \G0\ produce the observed column density of molecular oxygen.

\section{Conclusions}
\label{sec:conclusions}

We have presented the first deep (7.7 hr) {\it Herschel}-HIFI
observations of the \oofull\ line at 487~GHz towards a deeply embedded
Class~0 protostar, NGC~1333~IRAS~4A.
The results from the observations and models can be summarized as follows.

\begin{itemize}

\item No \oo\ emission is detected from the protostellar envelope,
  down to a 3$\sigma$ upper limit of \xoo\ $\lesssim$~6$\times$10$^{-9}$, the
  lowest \oo\ abundance limit toward a protostar to date. The O$_2$/CO
  limit is $\leq$6$\times$10$^{-3}$.

\item A full gas-grain chemical model coupled with the physical
  structure of the envelope is compared to our data consisting of two
  stages, a ``pre-collapse stage'' and ``protostellar stage''.  Best
  fits to the observed upper limit on the \oo\ line suggest a long
  pre-collapse stage ($\sim$0.7--1$\times$10$^{6}$~years), during which atomic oxygen is
  frozen out onto the dust grains and converted into water ice. Also,
  at least a fraction of NO must be converted to more complex nitrogen
  species in the ice.

\item The low \oo\ abundance in the gas and on the grains in the inner envelope implies
  that the material entering the disk is very poor in \oo.

\item A 4.5$\sigma$ tentative \oo\ detection is found at
  \vlsr\ =~8.0~\kms, which is interpreted as emission originating from
  the surrounding more extended NGC~1333 cloud. At this velocity, also a weak 
  NO line is seen.

\item Comparison with PDR models of \citet{Hollenbach09} suggests a
  high \G0\ of 300--650 in the surrounding cloud for the low inferred density of 
  <10$^4$ cm$^{-3}$. For the low \G0\ inferred at our position, 
  either the \hho\ ice photodesorption yield would need to be increased by a 
  factor of $\sim$2 or  combination of other minor changes in reaction rates 
  would be needed to reproduce the observed \NOO\ with a reasonable value of \G0.
\end{itemize}

\begin{acknowledgements}
UAY and astrochemistry in Leiden are supported by the Netherlands
Research School for Astronomy (NOVA), by a Spinoza grant and grant
614.001.008 from the Netherlands Organisation for Scientific Research
(NWO), and by the European Community's Seventh Framework Programme
FP7/2007-2013 under grant agreement 238258 (LASSIE) and 291141
(CHEMPLAN).  This work was carried out in part at the Jet Propulsion Laboratory, 
which is operated by the California Institute of Technology under contract with NASA. 
The authors are grateful to many funding agencies and the
HIFI-ICC staff, who has been contributing for the construction of
\textit{Herschel} and HIFI for many years. HIFI has been designed and
built by a consortium of institutes and university departments from
across Europe, Canada and the United States under the leadership of
SRON Netherlands Institute for Space Research, Groningen, The
Netherlands and with major contributions from Germany, France and the
US.  Consortium members are: Canada: CSA, U.Waterloo; France: CESR,
LAB, LERMA, IRAM; Germany: KOSMA, MPIfR, MPS; Ireland, NUI Maynooth;
Italy: ASI, IFSI-INAF, Osservatorio Astrofisico di Arcetri- INAF;
Netherlands: SRON, TUD; Poland: CAMK, CBK; Spain: Observatorio
Astron{\'o}mico Nacional (IGN), Centro de Astrobiolog{\'i}a
(CSIC-INTA). Sweden: Chalmers University of Technology - MC2, RSS $\&$
GARD; Onsala Space Observatory; Swedish National Space Board,
Stockholm University - Stockholm Observatory; Switzerland: ETH Zurich,
FHNW; USA: Caltech, NASA/JPL, NHSC.
\end{acknowledgements}

\bibliographystyle{aa}
\bibliography{bibdata}

\Online

\appendix

\section{HIFI \oo\ Spectrum}

\begin{figure*}[!t]
\begin{center}
\includegraphics[scale=0.64]{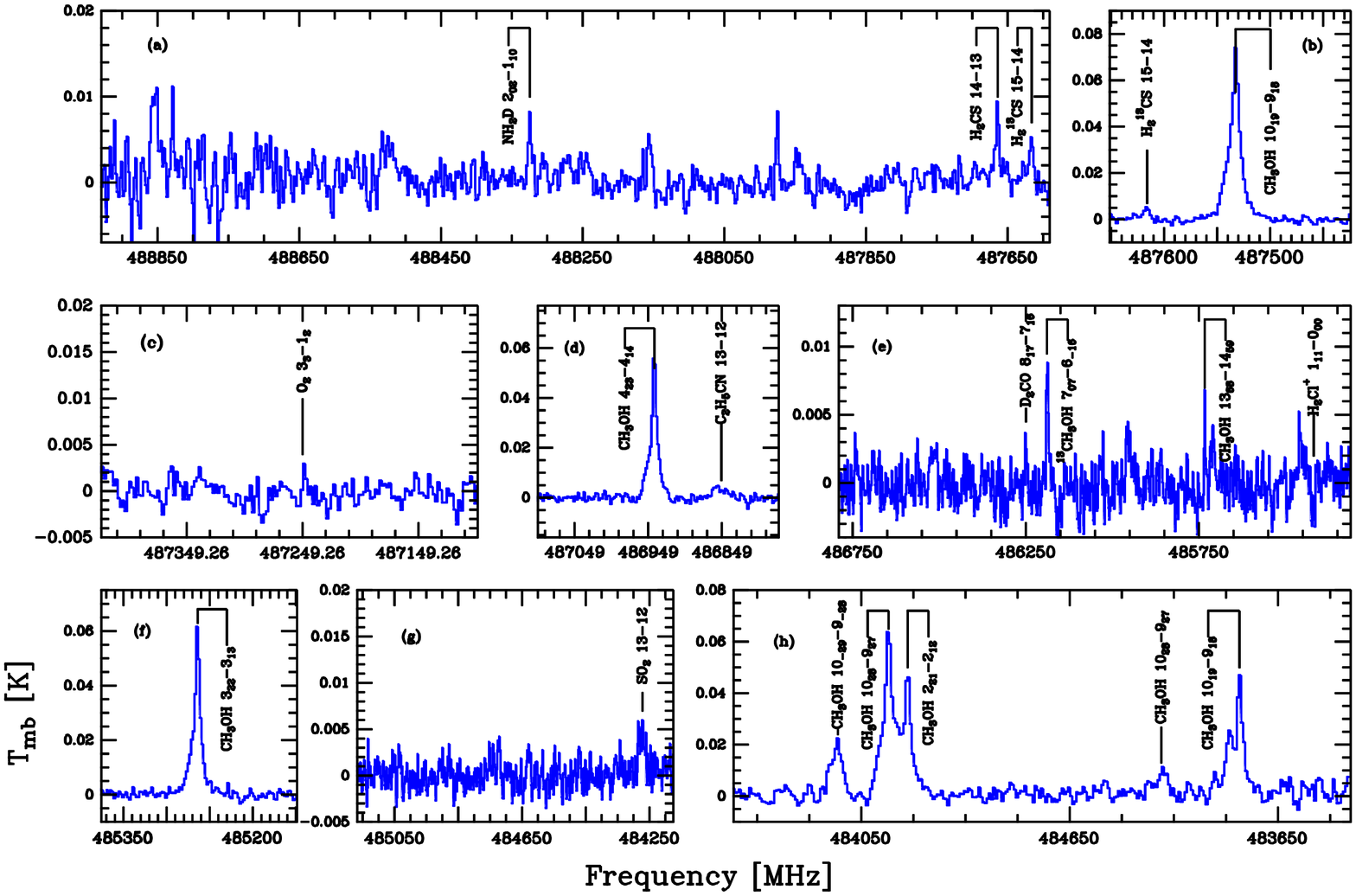}
\end{center}
\caption{\small{Spectrum of Fig. \ref{fig:iras4a487ghzfull} magnified over certain velocity ranges. In panel ({\it c}) the \oofull\ transition is shown. Identifications refer to the 7.0~\kms\ component.}}
\label{fig:iras4a487ghzpanels}
\end{figure*}

\begin{table*}[!th]
\caption{Overview of the other lines observed in the same spectrum. The level energies, Einstein A coefficients, and line frequencies are from the LAMDA, JPL and CDMS databases \citep{Schoier2005LAMDA, Pickett1998JPL, Muller2005CDMS}.}
\small
\begin{center}
\begin{tabular}{l r r r r r r r r r r}
\hline \hline
Mol. & Trans.& $E_\mathrm{u}/k_{\mathrm{B}}$ & $A_{\rm ul}$ & Frequency & $\int T_{\mathrm{MB}} \mathrm{d}V$ & $T_{\mathrm{peak}}$ & FWHM\\
 &  $J_{\mathrm{u}}$-$J_{\mathrm{l}}$ &  [K] & [s$^{-1}$] & [GHz]  & [mK~km~s$^{-1}$] & [mK] & [km s$^{-1}$] \\
\hline
\noalign{\smallskip}
NH$_2$D         & 2$_{02}$--1$_{10}$  & 47.2  & 1.36$\times$10$^{-4}$ & 488.323810 & 25 & 10\phantom{0}  &  3.4 \\
H$_2$CS         & 14--13              & 188.8 & 1.76$\times$10$^{-3}$ & 487.663321 & 28 & 9  &  3.3 \\
H$_2^{13}$CS    & 15--14              & 200.5 & 1.77$\times$10$^{-3}$ & 487.615288 & 22 & 5  &  3.9 \\
CH$_3$OH        & 10$_{19}$--9$_{18}$ & 143.3 & 5.15$\times$10$^{-4}$ & 487.531887 & 580 & 80 & 8.2\\
O$_2$           & 3$_{3}$--1$_{2}$    & 26.38 & 8.66$\times$10$^{-9}$ & 487.249264 & 7 & \dots & \dots \\
CH$_3$OH        & 4$_{23}$--4$_{14}$  & 60.9  & 5.45$\times$10$^{-4}$ & 486.940837 & 390 & 60 & 7.2 \\
C$_2$H$_5$CN    & 13--12              & 66.9  & 1.00$\times$10$^{-6}$ & 486.849912 & 36 & 5 & 7.3\\
D$_2$CO         & 8$_{17}$--7$_{16}$  & 111.0 & 3.36$\times$10$^{-3}$ & 486.248662 & 9 & 7 & 1.1\\
$^{13}$CH$_3$OH & 7$_{07}$--6$_{16}$  & 76.5  & 3.02$\times$10$^{-4}$ & 486.188242 & 39 & 10 & 3.9\\
CH$_3$OH        & 13$_{68}$--14$_{59}$& 404.8 & 1.16$\times$10$^{-4}$ & 485.732280 & 12 & 10 & 1.4\\
H$_{2}$Cl$^{+}$ & 1$_{11}$--0$_{00}$  & \dots & \dots                 & 485.420796 & $-$5 & $-$6 & 0.8 \\ 
CH$_3$OH        & 3$_{22}$--3$_{13}$  & 51.6  & 5.02$\times$10$^{-4}$ & 485.263263 & 401 & 71 & 7.3 \\
SO$_2$          & 13--12              &105.8  & 5.42$\times$10$^{-4}$ & 484.270879 & 26 & 6 & 1.1 \\
CH$_3$OH        & 10$_{-29}$--9$_{-28}$&153.6 & 4.88$\times$10$^{-4}$ & 484.071775 & 170 & 20 & 8.4 \\
CH$_3$OH        & 10$_{28}$--9$_{27}$ & 150.0 & 4.83$\times$10$^{-4}$ & 484.023168 & 490 & 70 & 8.2 \\
CH$_3$OH        & 2$_{21}$--2$_{12}$  & 44.7  & 3.99$\times$10$^{-4}$ & 484.004740 & 280 & 50 & 4.9 \\
CH$_3$OH        & 10$_{28}$--9$_{27}$ & 165.4 & 4.90$\times$10$^{-4}$ & 483.761387 & 70 & 10 & 5.8 \\
CH$_3$OH        & 10$_{19}$--9$_{18}$ & 148.7 & 5.13$\times$10$^{-4}$ & 483.686308 & 210 & 50 & 4.2 \\
\noalign{\smallskip}
\hline
\end{tabular}
\end{center}
\tablefoot{\rms\ is 1.3~mK and in 0.35~\kms\ bin. Identifications refer to the 7.0~\kms\ component.
}
\label{tbl:overviewobsHIFI}
\end{table*}

\section{RADEX Calculations for \ceio}
\label{sec:radexc18o}

\begin{figure*}[!th]
\begin{center}
\includegraphics[scale=0.25]{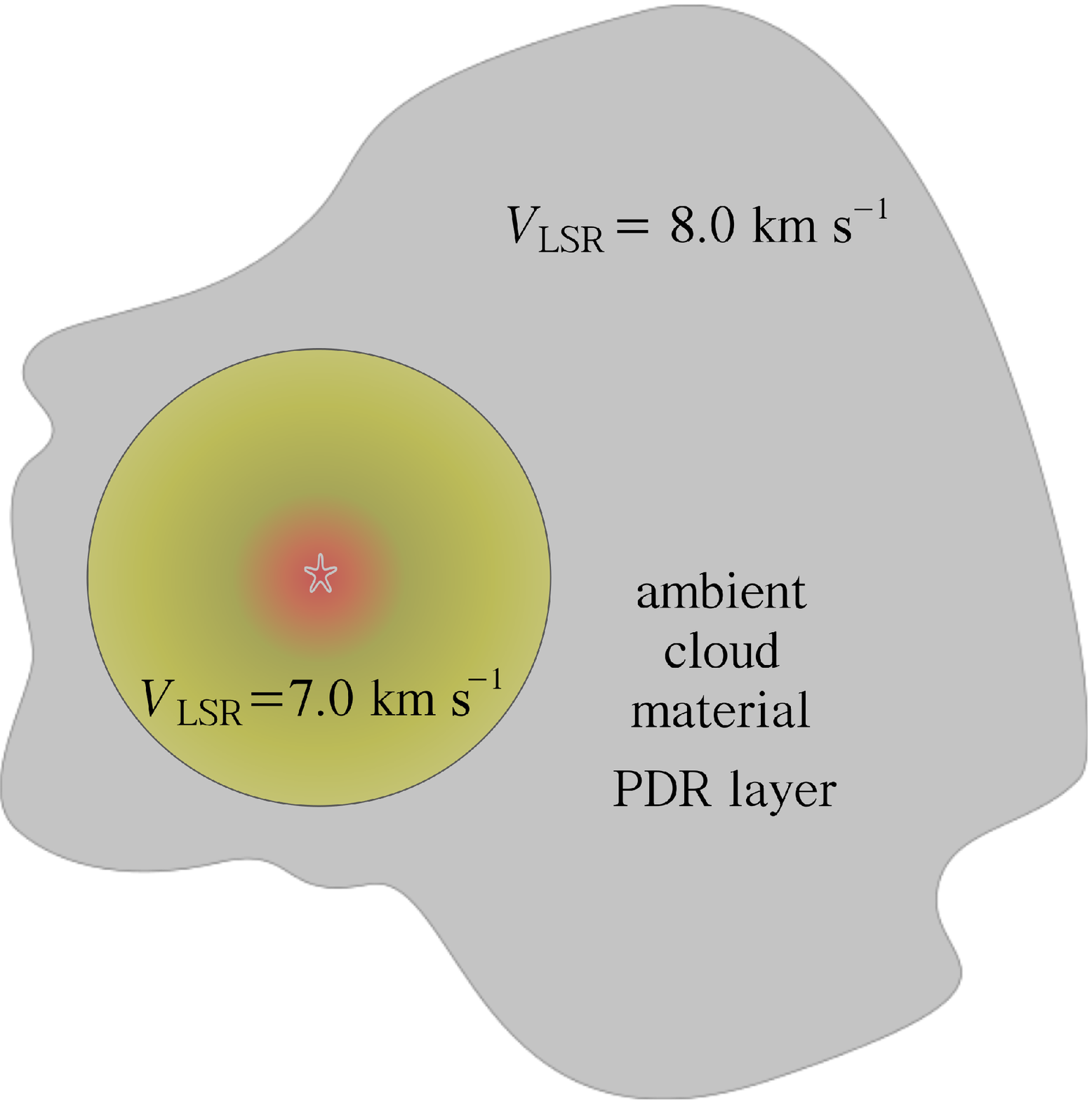}
\includegraphics[scale=0.50]{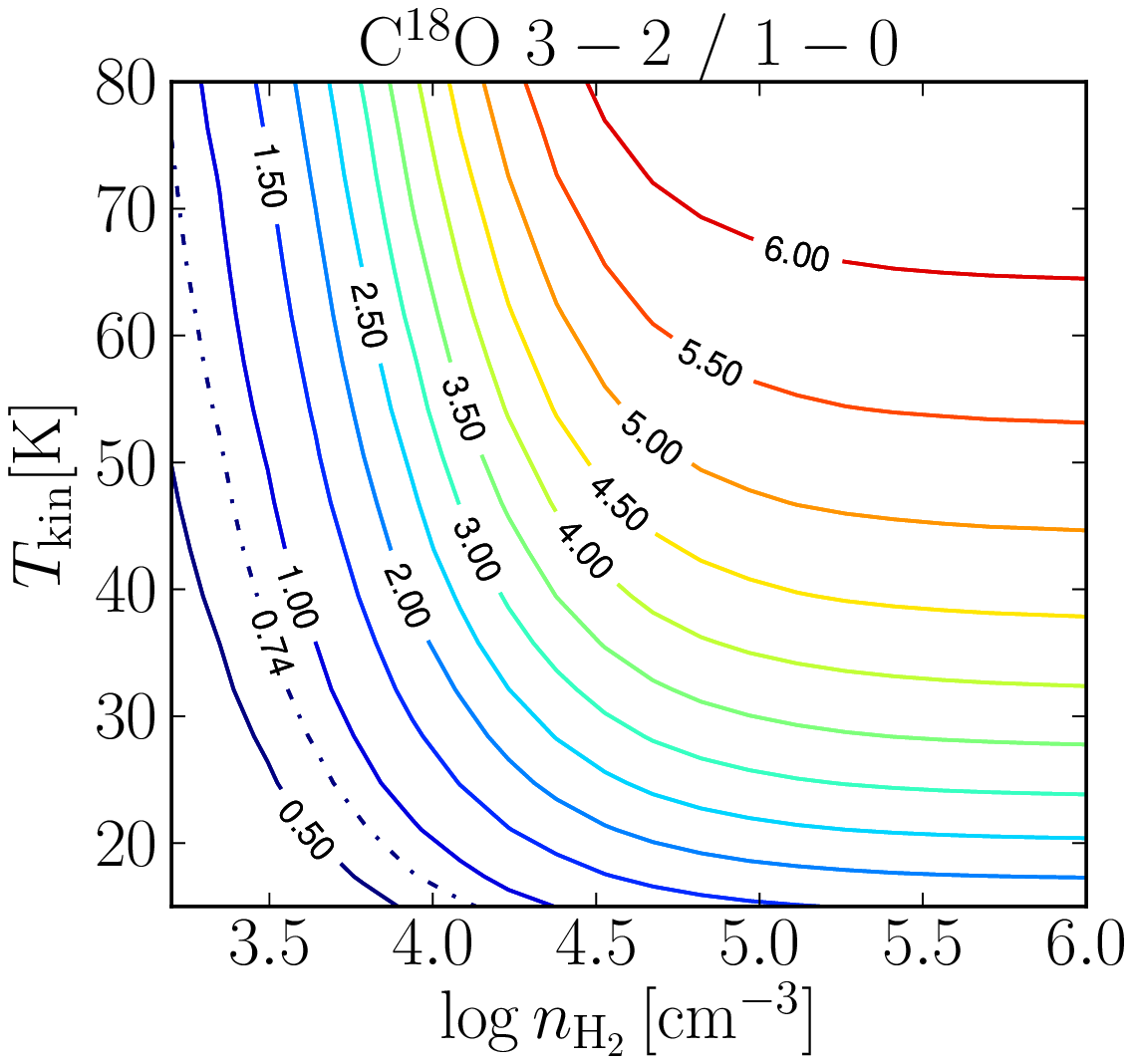}
\end{center}
\caption{\footnotesize{{\it Left:} Schematic cartoon showing
      the scenario of \oo\ emission originating from the surrounding
      cloud. {\it Right:} Integrated intensity ratios calculated with
    RADEX, as function of temperature and density, for a \ceio\ column
    density of 5$\times$10$^{14}$~\cmtwo\ (optically thin
    conditions). The \ceio\ 3--2/1--0 ratio is relevant for the
    surrounding NGC~1333 cloud, which is traced by the 8.0~\kms\
    component.  Dash-dotted lines indicate the observed ratio of
    \ceio\ 3--2/1--0=0.74 for the \vlsr=8.0 \kms\ component.}}
\label{fig:radex}
\end{figure*}

The integrated intensity ratio of \ceio\ 3--2/1--0 is equal to 0.74 for
the 8.0~\kms\ component.  This ratio can be analysed using the
RADEX non-LTE excitation and radiative transfer program
\citep{vanderTak07} to constrain the physical parameters.  Figure
\ref{fig:radex} presents the integrated intensity ratios as function
of temperature and density, obtained for optically thin
conditions. The observed ratio is indicated in dash-dotted lines.

\end{document}